\documentclass[aps,prb,twocolumn]{revtex4}
\usepackage{color}
\usepackage{float}
\usepackage{amsmath}
\usepackage{amssymb}
\usepackage{graphicx}
\begin{document}
\title{Two-electron n-p double quantum dots in carbon nanotubes}
\author{E. N. Osika}
\author{B. Szafran}
\affiliation{AGH University of Science and Technology, Faculty of Physics and Applied Computer Science,\\
al. Mickiewicza 30, 30-059 Krak\'ow, Poland}
\begin{abstract}
We consider electron states in n-p double quantum dots defined in a semiconducting carbon nanotube (CNT) by an external potential.
We describe formation of extended single-electron orbitals originating from the conduction and valence bands confined in a minimum and a maximum of the external potential, respectively.
We solve the problem of a confined electron pair using an exact diagonalization method within the tight-binding approach, which allows for a straightforward treatment of the conduction and valence band states, keeping an exact account for the intervalley scattering mediated by the atomic defects and the electron-electron interaction.
The exchange interaction -- which in the unipolar double dots is nearly independent of the axial magnetic field ($B$) and forms singlet-like and triplet-like
states -- in the n-p system appears only for selected states and narrow intervals of $B$.
In particular the ground-state energy level of a n-p double dot is not split by the exchange interaction and remains four-fold degenerate at zero magnetic field
also for a strong tunnel coupling between the dots.
\end{abstract}
\maketitle

\section{Introduction}

Due to the absence of the hyperfine interaction the graphene-based \cite{graphene} materials are an attractive
medium for spin control and manipulation.
In semiconductor carbon nanotubes \cite{cnts,lairdrev} formation of the energy gap prevents the Klein tunneling \cite{kt} and allows for confinement of charge carriers in quantum dots formed by external voltages.
The transport spectroscopy \cite{kuemmeth,jesp,soc1} experiments resolve the signatures of the  spin-orbit coupling that appears \cite{ando,brataas,chico,valle,Bulaev,klino1,flensberg,str2} with folding of the graphene plane into a nanotube.
The spin-orbit interaction induces formation of spin-valley states \cite{ando,brataas,chico,valle,Bulaev,klino1,flensberg,str2} through coupling of the orbital magnetic moments with spin.
The effects of spin and valley dynamics are monitored in the electric dipole spin-valley resonance experiments \cite{pei,pei2,np} by lifting the valley and / or
spin blockade \cite{pal} of the current flow through a pair of quantum dots connected in series.

The pair of quantum dots
confining localized electron spins \cite{loss} is the basic element of the quantum information processing circuitry.
The effective spin exchange interaction that splits the
singlet and triplet energy levels is a necessary prerequisite for construction of a universal quantum gate \cite{loss}.
In single \cite{jesp,wunsch,rontani2,steele} and double CNT quantum dots \cite{stecher,weiss,rey,pei} the coupling of the spin and valley degrees of freedom  results
in formation of singlet-like and triplet-like  states of the electron pair. These states are no longer spin eigenstates but they still possess a definite symmetry of the spatial wave function with respect to the electron interchange.

\begin{figure}[htbp]
\includegraphics[width=7cm]{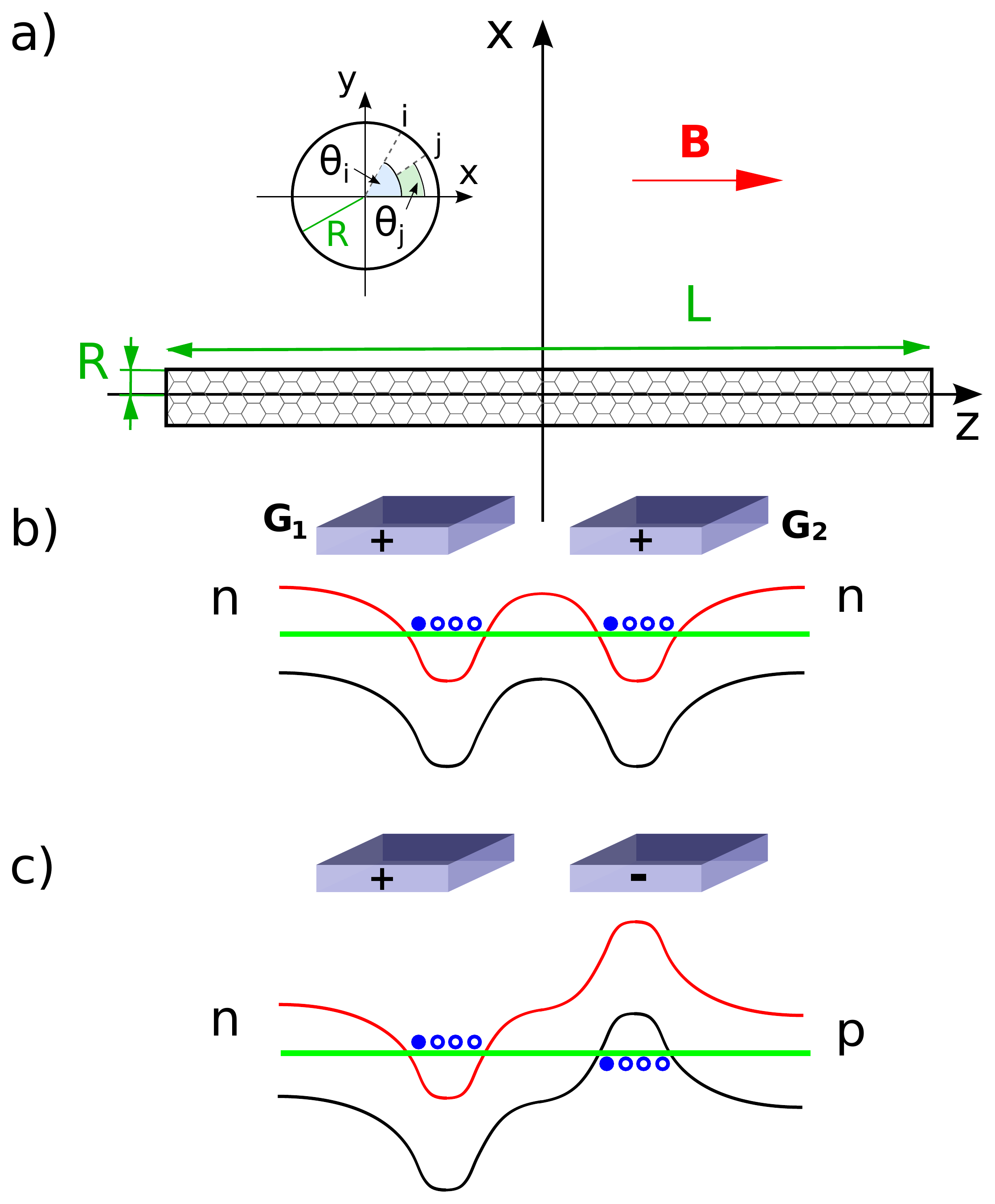}
\caption{(a) Schematics of the considered system. The external magnetic field is oriented
along the axis $z$ of the zigzag CNT of radius $R$ and length $L$. The inset explains the
angles used for the definition of the interatom hopping elements of the tight-binding Hamiltonian
in presence of the spin-orbit interaction.
We study the system of a double n-n dot (b) or n-p dot (c) induced by external voltages.
The discussed states correspond to a single electron per dot: occupying
one of the four-fold degenerate confined energy levels. The green line indicates the Fermi energy.
In the considered low-energy states the electrons occupy mostly separated quantum dots.}
\label{schemat}
\end{figure}

The graphene is an ambipolar material and the external potentials  easily sweep the conduction and valence band extrema above or below the Fermi energy \cite{zeb,pei}. In the spin-valley resonance experiments \cite{pei,pei2}
the double quantum dot is set in a n-p configuration for which the Pauli
blockade is most pronounced.
In the present paper we describe formation of electron orbitals extended over the n-p double quantum dot. Next, we study the spin-valley structure of the two-electron states
with a single electron in the four-fold degenerate confined state per dot (see Fig. \ref{schemat}),
which in experimental papers \cite{pei,pei2} is addressed as (3h,1e) -- a charge configuration with three holes in one quantum dot and a single electron in the other.
The two-electron system in the n-p double dot is usually considered similar \cite{pei} to the electron-pair in the n-n double dots \cite{stecher,weiss,rey}.
Here, we demonstrate that the electronic structure of the double dot n-p system differs in a few elementary aspects:
 {\it i)} the energy level splitting by the spin-exchange interaction is missing in the two-electron ground-state which is four-fold degenerate also
 when the tunnel coupling between the dots is strong;
{\it ii)} the splitting resulting from  the exchange energy is found only in the excited part of the spectrum and for a limited range of magnetic fields; {\it iii)} formation
of singlet-like and triplet-like spatial orbitals appears only within avoided crossings induced by the external electric or magnetic fields.
We indicate that these features result from an
 opposite electron circulation in the conductance and valence bands for a given valley.
Formation of extended orbitals in the n-p double dots
in presence of the spin-orbit coupling introduces a dependence of the electron distribution on the spin and valley,
which produces a fine structure of the two-electron spectrum at low $B$. For completeness we include a brief tight-binding analysis of the n-n system, which has been considered
in the continuum approximation in Refs. \cite{stecher,weiss}.

The present study is based on the exact diagonalization approach using the tight-binding method that allows for a consistent description of conduction and valence band states, intervalley mixing  due to the atomic disorder and the short-range component of the Coulomb interaction  \cite{umklapp}, and does not require an additional parametrization.
The intervalley scattering due to the electron-electron interaction is usually neglected by  effective mass theories \cite{ando1}. The tight binding approach at the configuration-interaction level \cite{potasz,potasz2} accounts for all the intervalley scattering processes which result from the electron-electron interaction, including the backward and umklapp scattering \cite{umklapp}. Inclusion of the intervalley scattering effects  to the low-energy theories is possible \cite{mayrhoer,rontani} but far from straightforward. Finally, the tight-binding approach accounts for even large
modulation of the external potential defining the CNTs, which is not necessarily the case for the low-energy continuum approximations.


\begin{figure}[htbp]
\includegraphics[width=7cm]{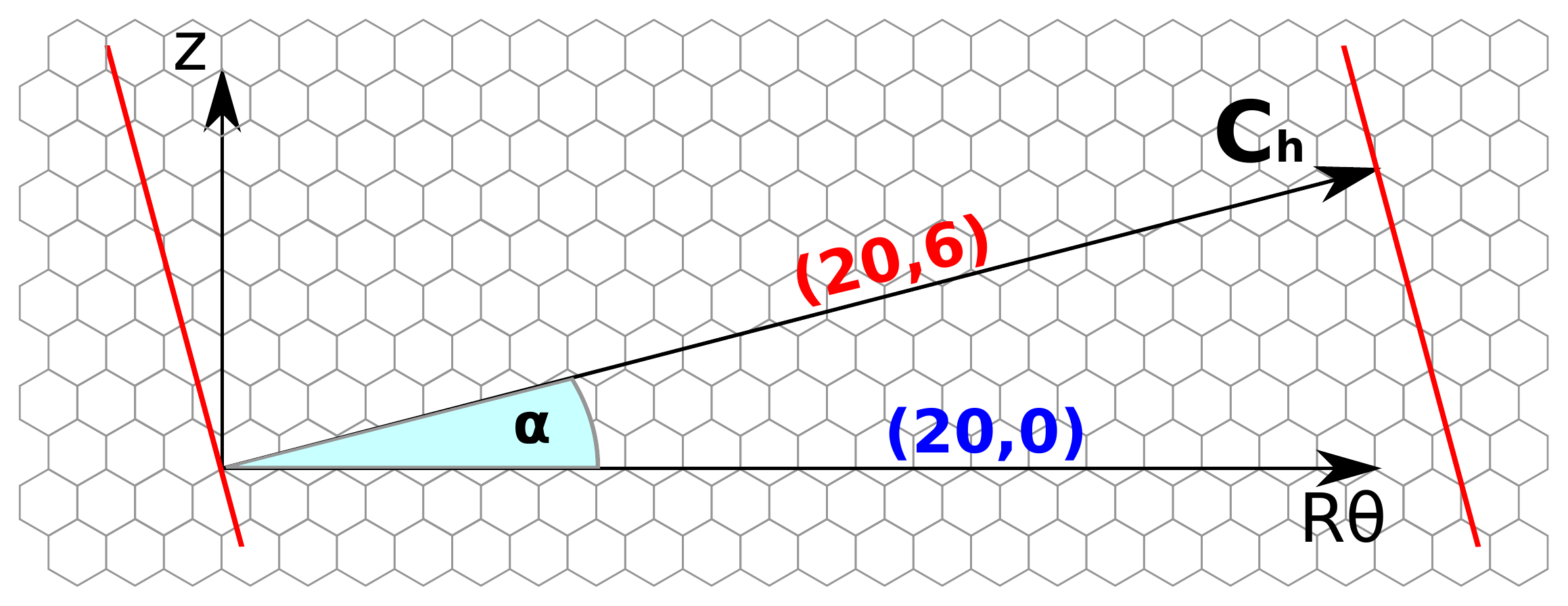}
\caption{Schematics of the CNT folding for the chiral vector $C_h=(20,0)$ (a zigzag CNT)
and $C_h=(20,6)$ which are considered in this work. }
\label{schematchiral}
\end{figure}

\section{single-electron states: theory}

We consider a semiconducting nanotube of length $L$ and radius $R$ [see Fig. \ref{schemat}(a)].
Most of the results are obtained for a zigzag CNT of length $L=53.1$ nm with 20 atoms along the circumference (diameter $2R=1.56$ nm).
The properties  of the low-energy two-electron  states in the double dots as determined for the zigzag CNT [$C_h=(20,0)$] are reproduced for any semiconducting CNT. For demonstration we provide below (Section \ref{chirals}) also the results for $C_h=(20,6)$ CNT chirality  (see Fig. \ref{schematchiral}).

We use the tight-binding Hamiltonian of the form
\begin{multline}
H=\sum_{\{i,j,\sigma,\sigma'\}}(c_{i\sigma}^\dagger \cdot t_{ij}^{\sigma\sigma'} \cdot  c_{j\sigma'}+h.c.)\label{eqh}\\
+\sum_{i,\sigma,\sigma'}c_{i\sigma}^\dagger  \cdot  \left(W_{QD}({\bf r}_i)+\frac{g\mu_b}{2}  \boldsymbol{ \sigma}\cdot {\bf B}\right) \cdot c_{i\sigma'},
\end{multline}
where the first summation runs over $p_z$ spin-orbitals of nearest neighbor pair of atoms, $c_{i\sigma}^\dagger$ $(c_{i\sigma})$ is the particle creation (annihilation) operator at ion $i$ with spin $\sigma$ in $z$ direction, and  $t_{ij}^{\sigma\sigma'}$ is the hopping parameter.
The second summation in Eq. (\ref{eqh})   accounts
for the external potential and the Zeeman interaction. In Eq. (1) $g=2$ is the Land\`e factor
and  $\boldsymbol{\sigma}$ stands for the vector of Pauli matrices.
The external magnetic field ${\bf B}=(0,0,B)$ is applied along the axis of the CNT.

The energy gap of the considered CNTs allows for electrostatic confinement of the carriers.
The quantum dot confinement is induced by external potentials
modeled by a sum of Gaussian functions:
\begin{equation}
W_{QD}({\bf r})=V_l \exp(-(z+z_s)^2/d^2)+ V_r\exp(-(z-z_s)^2/d^2)\, ,
\end{equation}
where $z_s$ is the shift of the dots from the center of the CNT ($z=0$),
$V_l$ and $V_r$ are potentials of the left and the right dot, respectively.

The paper is focused on the states with a single electron per quantum dot [cf. Fig. \ref{schemat}(b,c)].
For separated electrons the details of the single-dot potential are of secondary importance
for the qualitative properties of the system as long as the tunnel coupling between the dots is present.
Most of the discussion is carried for small quantum dots
with $2d=4.4$ nm, with the shift between their centers $2z_s=10$ nm.
For these small quantum dots the single-electron energy level spacing is large ($\simeq 100$ meV)
which is useful for analysis of the properties of the exchange interaction, since a limited
number of multiplets contribute to the two-electron wave functions.
Nevertheless, the single-particle level spacings in CNT quantum dots is of the order of a few meV, up to 10 meV at most \cite{soc1,str2}.
In order to demonstrate that the identified properties of the n-p system are qualitatively independent
of the size of the dots we provide in Section (\ref{sectionlarge}) also the results for larger QDs.

The hopping parameters $t_{ij}^{\sigma\sigma'}$ between the nearest neighbor spin-orbitals -- including the curvature induced spin-orbit coupling \cite{ando,chico,valle} -- are introduced in the following form \cite{ando,valle}
\begin{eqnarray}
t_{ij}^{\uparrow\uparrow}&=&V_{pp}^{\pi}\cos(\theta_{i}-\theta_{j})\nonumber \\
&-&(V_{pp}^{\sigma}-V_{pp}^{\pi})\frac{r^{2}}{a_{C}^{2}}[\cos(\theta_{i}-\theta_{j})-1]^{2}+
\nonumber \\
&+&2i\delta\big\{ V_{pp}^{\pi}\sin(\theta_{i}-\theta_{j})+ \nonumber \\&&(V_{pp}^{\sigma}-V_{pp}^{\pi})\frac{r^{2}}{a_{C}^{2}}\sin(\theta_{i}-\theta_{j})[1-\cos(\theta_{i}-\theta_{j})]\big\} \nonumber \\&=&{t_{ij}^{\downarrow\downarrow}}^*
\end{eqnarray}
\begin{eqnarray}
t_{ij}^{\uparrow\downarrow}&=&-\delta(e^{-i\theta_{j}}+e^{-i\theta_{i}})(V_{pp}^{\sigma}-V_{pp}^{\pi})\frac{rZ_{ji}}{a_{C}^{2}}[\cos(\theta_{i}-\theta_{j})-1] \nonumber
\\ &=&-{t_{ij}^{\downarrow\uparrow}}^*\label{spino}
\end{eqnarray}
where  $V_{pp}^{\pi}=-2.66\,$eV, $V_{pp}^{\sigma}=6.38\,$eV,\cite{tomanek}
$a_C=0.142$ nm is the nearest neighbor distance,
$\theta_{i}$ indicates the localization angle of atom $i$ in the $(x,y)$ plane [see the inset to Fig. 1(a)],
and  $Z_{ji}=Z_{j}-Z_{i}$ is the distance between atoms $i$ and $j$ along the CNT axis.
The SO coupling  parameter is taken $\delta=0.003$ \cite{ando,valle} unless explicitly stated otherwise.

Orbital effects of the external magnetic field are introduced by Peierls phase shifts
$t_{ij}^{\sigma\sigma'}\rightarrow t_{ij}^{\sigma\sigma'}e^{i2\pi (e/h)\int_{{\bf r}_{i}}^{{\bf r}_{j}}\boldsymbol{A}\cdot\boldsymbol{dl}}.$
We apply the Landau  gauge ${\boldsymbol{A}}=(0,Bx,0)$.

In the following we refer to electron currents circulating along the circumference of the nanotube. In the tight-binding model the operator of the probability current \cite{current} flowing along
the $\pi$  bonds between $k$-th and $l$-th neighbor ion spin-orbitals is given by the formula
\begin{equation}
 J_{kl}^{\sigma\sigma'}=\frac{i}{\hbar}(c_{k\sigma}^\dagger \cdot t_{kl}^{\sigma\sigma'} \cdot  c_{l\sigma'}-h.c.),
\end{equation}
which accounts for the spin-precession due to the spin-orbit interaction.
In the following discussion we refer to the dominating, i.e. the spin-conserving components of the current.

\section{single-electron states: Results}

\subsection{Separate $n$ and $p$ quantum dots}

\begin{figure}[htbp]
\includegraphics[width=7.8cm]{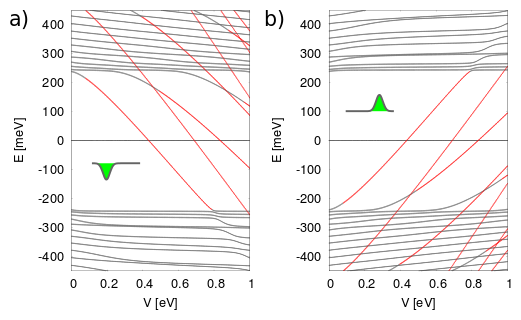}
\caption{a) Energy spectrum for a CNT with a local potential minimum (a) and maximum (b) introduced by an external potential
as functions of the depth (a) and height (b) of the Gaussian potential well (a) and barrier (b). With the red lines we plotted
the energy levels that correspond to electron localization inside the Gaussian (within the central segment of length $2d$) by at least 50\%.}\label{dg}
\end{figure}

\begin{figure}[htbp]
\includegraphics[width=7.2cm]{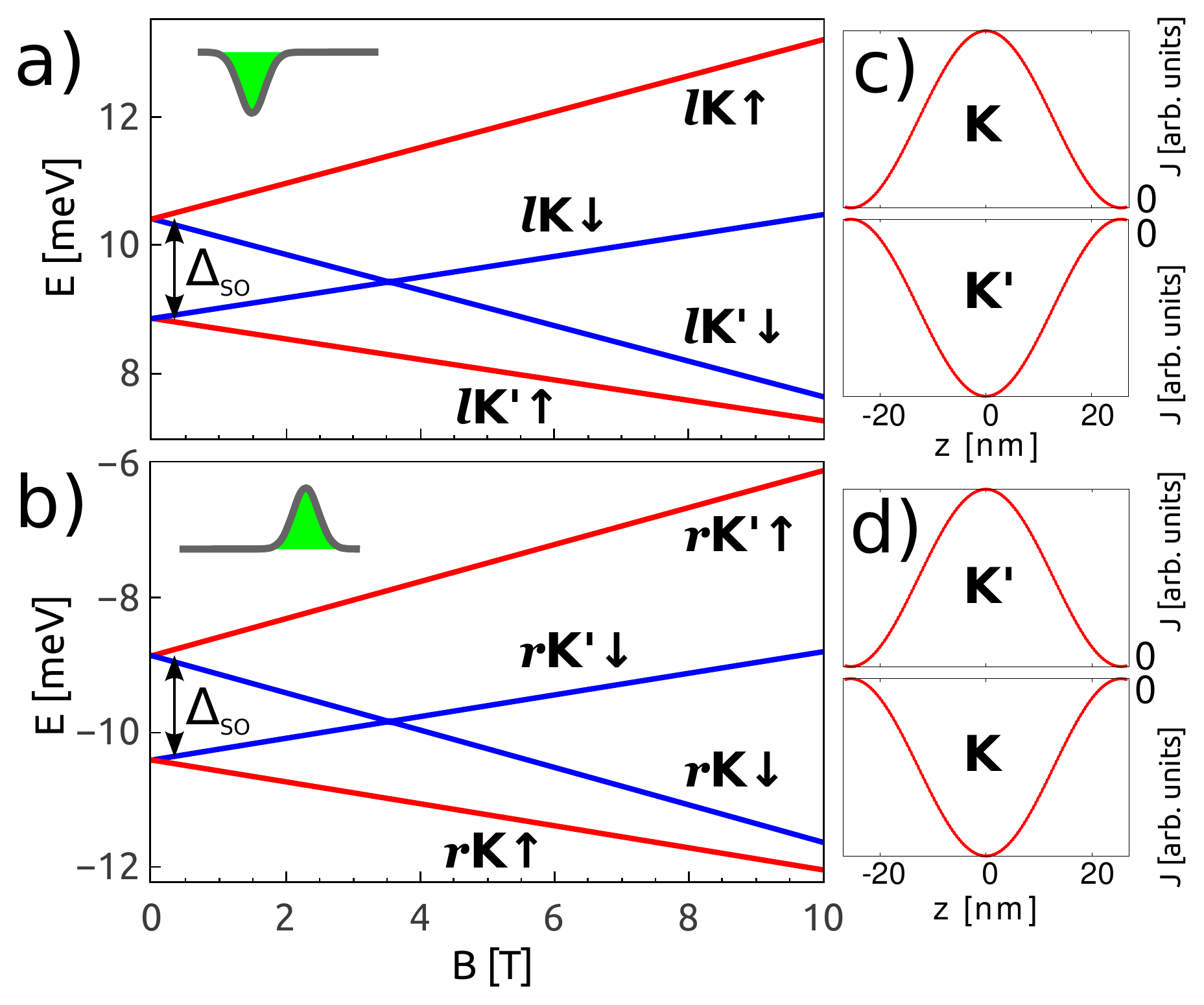}
\caption{Energy levels for the single-electron states localized inside a single separate n-quantum dot (a) or p-quantum dot (b) as function of the external magnetic field for $V=\pm 0.42$ eV.
The energy levels are labeled by valley $K/K'$, spin $\uparrow\downarrow$, $l$ and $r$ denote the left / right dot. In (c) [(d)] we plotted the circumferential component of the electron current calculated at
$y=0$ for the lowest (highest) energy states of the conduction (valence) bands for $V=0$ in the absence of the spin-orbit coupling.
}\label{wgd}
\end{figure}
Figure \ref{dg} shows the energy spectrum for a single external Gaussian potential introduced as a minimum [n-type quantum dot, Fig. \ref{dg}(a) for $V_l<0$, $V_r=0$] or a maximum [p-type quantum dot, Fig. \ref{dg}(b) for $V_l=0,V_r>0$]
inside the carbon nanotube. The energy levels plotted in red correspond to states localized inside the n-  [Fig. \ref{dg}(a)]  or p-type quantum dot [Fig. \ref{dg}(b)].
With the external potential that is introduced to the CNT, the energy spectrum is no longer symmetric with respect to the zero energy.
The  spectrum for the n-type dot [Fig. \ref{dg}(a)] with the localized states evolving from the conduction band
is opposite to the spectrum for the p-dot [Fig. \ref{dg}(b)] with the localized states that evolve from the valence band.
All the localized energy levels are nearly four-fold degenerate with respect to the valley and spin -- the SO coupling energy $\Delta_{SO}$
is below the resolution of this plot.

Figure \ref{wgd} shows the calculated energy spectrum as a function of the external magnetic field for the single-electron states localized inside the n-type  [Fig. \ref{wgd}(a)]
and p-type dots [Fig. \ref{wgd}(b)] for $V=\pm 0.42$ eV.
In the n-type dot for $B=0$ one finds a Kramers doublet ($K'\uparrow$, $K\downarrow$) ground state
split by the spin-orbit interaction from  higher-energy doublet ($K'\downarrow$, $K\uparrow$).\cite{uwaga}
 The spin-orbit splitting of the energy levels of Fig. \ref{wgd} is $\Delta_{SO}$= 1.55 meV.
The degenerate $K$ and $K'$ states have an opposite orientation of the current circulation around the
axis of the nanotube \cite{minot}. For illustration we plotted the circumferential component of the current calculated \cite{current}
for $y=0$, and $V_l=V_r=0$ in the lowest state of the conduction band.
The conduction band low-energy $K'$ states that we deal with
produce orbital magnetic moment which is oriented in the $z$ direction, i.e. parallel to the external magnetic field.
The electron circulation in the $K$ states of conduction band is opposite [Fig. \ref{wgd}(c)].
Formation of the degenerate pairs of spin-valley energy levels $(K'\uparrow,K\downarrow)$ and $(K'\downarrow,K\uparrow)$ results from the curvature-induced spin-orbit coupling \cite{ando,brataas,klino1,valle,Bulaev}.
For the electrons localized inside the p-type dot [Fig. \ref{wgd}(b)]-- filling the states of the valence band  -- the spin-orbit coupling
produces a lower-energy doublet ($K\uparrow$, $K'\downarrow$) and a higher-energy one ($K\downarrow$, $K'\uparrow$).
The orbital moments for a given valley are opposite in the states of conduction
and valence bands \cite{minot} -- cf. the calculated electron current orientation in Fig. \ref{wgd}(c,d) --
thus in the lower-energy Kramers doublets of the p- and n-type dots the valleys are interchanged.
As we discuss below, this fact has a pronounced influence on the properties of the two-electron states for the n-p double quantum dots.

\subsection{Double quantum dots}
\label{smol}
\begin{figure}[htbp]
\includegraphics[width=7.8cm]{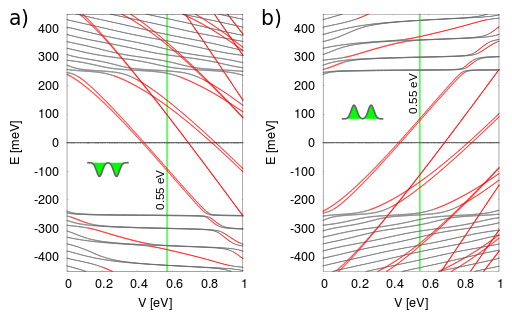}
\caption{Energy levels for a system of n-n (a) and p-p double dots (b) as a function of the depth / height of the Gaussian quantum dots / antidots.}\label{2kkd}
\end{figure}

Figure \ref{2kkd}(a) shows the energy spectrum for a double unipolar n-n quantum dot
in the (1e,1e) charge configuration as a function of the depth of the Gaussian quantum dots. For comparison in Fig. \ref{2kkd}(b) the energy spectrum for the p-p dot in the (3h,3h) charge states is shown.
For the double n-n [Fig. \ref{2kkd}(a)] and p-p dots [Fig. \ref{2kkd}(b)] we observe that the energy levels move in pairs with $V$. The pairs  correspond to  bonding and antibonding
orbitals extended over both the quantum dots \cite{weiss}. Each energy level within the pair is nearly 4-fold degenerate with respect to the valley and the spin.
The energy splitting between bonding and antibonding orbital $\Delta_{ba}$ is a few times larger than the spin-orbit splitting $\Delta_{SO}$ between
Kramers doublets within each of the orbitals (i.e. for $V=0.55$ eV for the lowest localized n-n states
$\Delta_{SO}\approx1.4$ meV and  $\Delta_{ba}\approx7.5$ meV).

\begin{figure}[htbp]
\includegraphics[width=7.2cm]{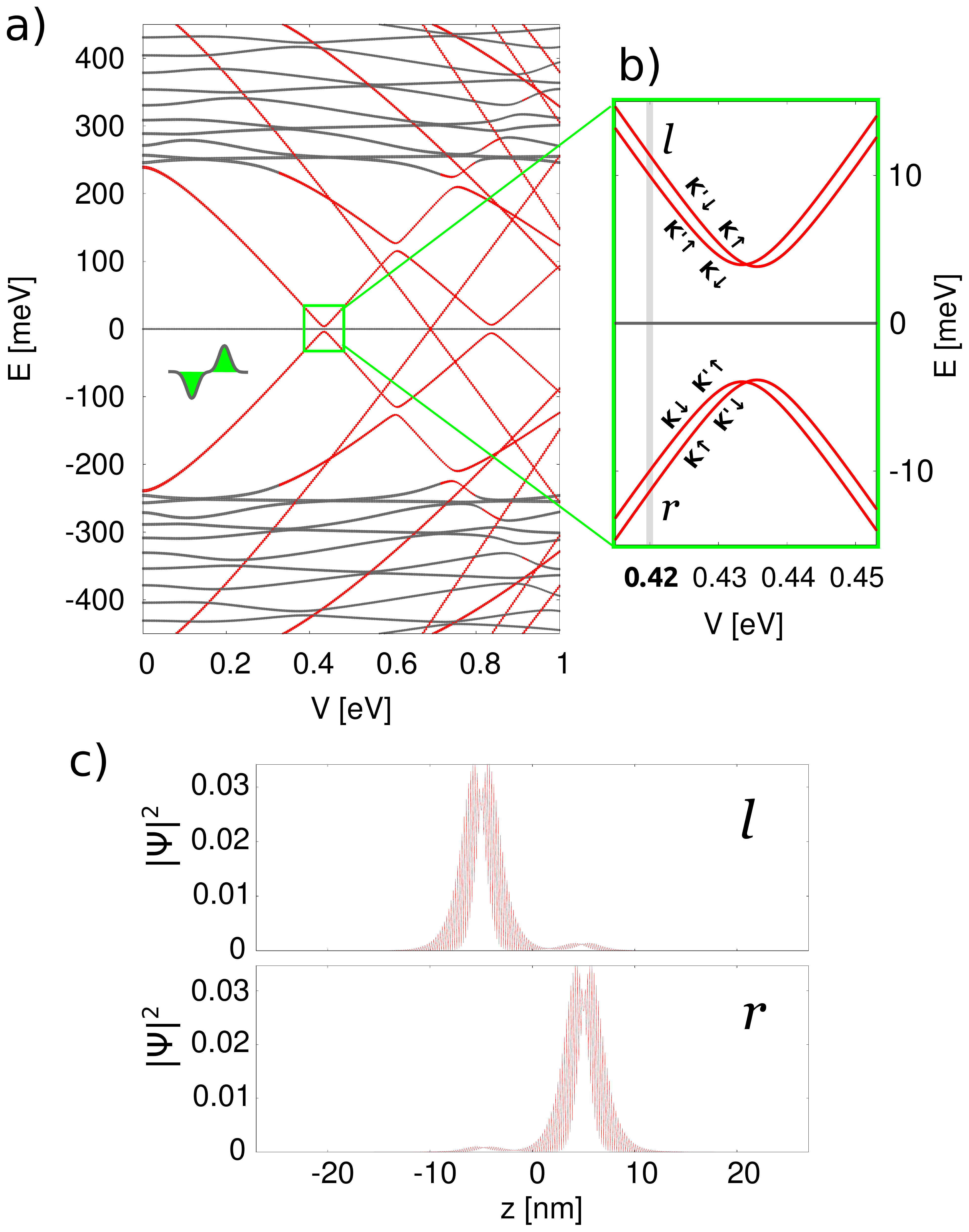}
\caption{(a) Energy spectrum for a n-p double dot as a function of the depth/height of the Gaussian potential $-V_l=V_r=V$. (b) Zoom at the avoided crossing of valence and conduction
band states  near the neutrality point. (c) Charge densities integrated along the circumference of the CNT for $V=0.42$ eV. The rapid oscillation results from contributions of A and B sublattices
which are both smooth but shifted one with respect to the other. }\label{np}
\end{figure}

\begin{figure}[htbp]
\includegraphics[width=7.5cm]{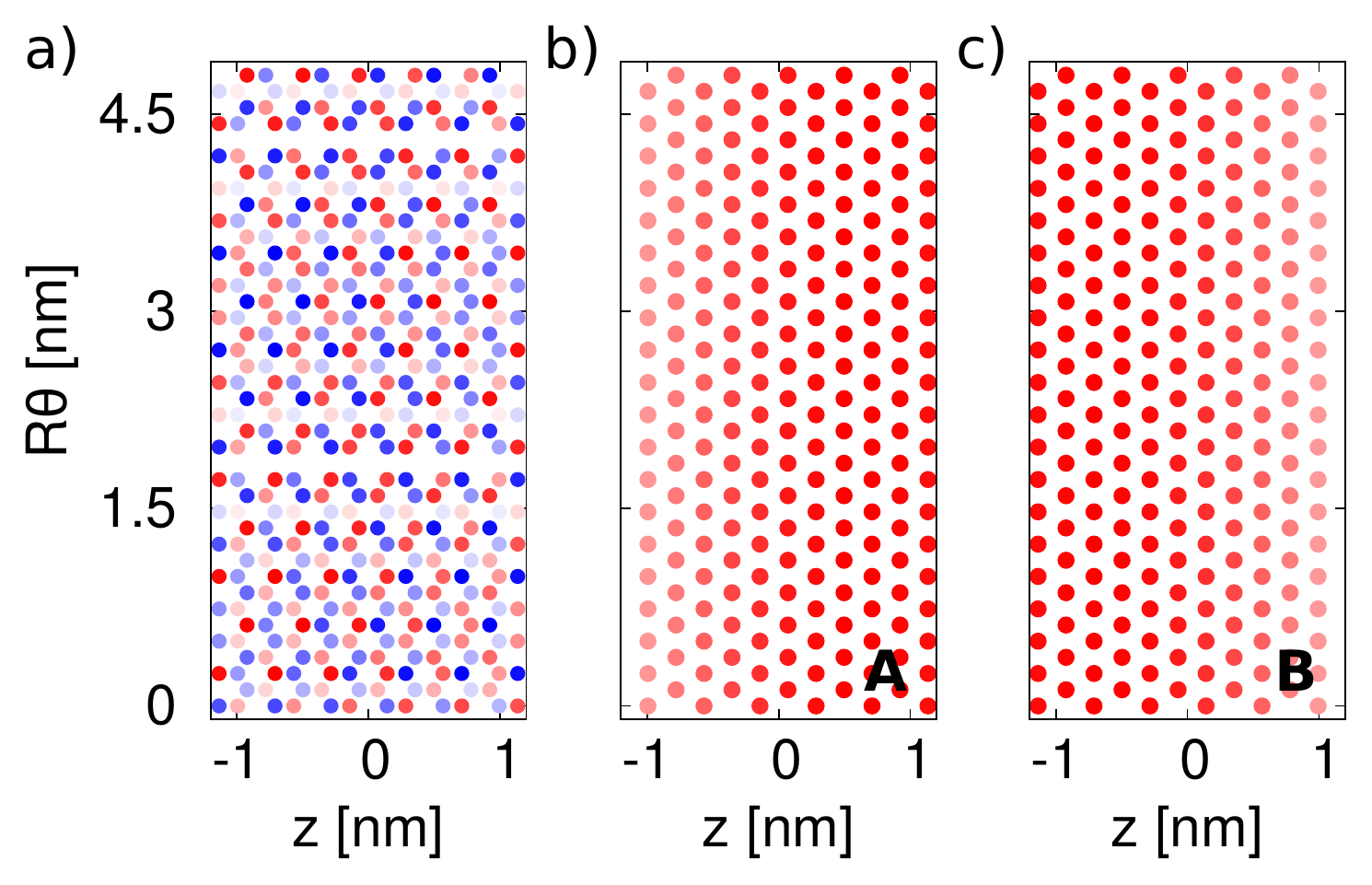}
\caption{The lowest-energy confined $K'\uparrow$ state in a single n-type quantum dot. A short fragment of the nanotube within the dot is considered.
 (a) Real part of the  majority spin component of $\Psi_{K'\uparrow}$ wave functions. The values that are plotted in red and blue correspond
 to positive and negative values, respectively. (b),(c) Real part of the envelope $u({\bf r})$ [see Eq. \ref{env}] on  sublattices A and B, respectively.
 }\label{ff}
\end{figure}

For the n-p double quantum dot  [Fig. \ref{np}(b)] the energy levels originating from the conduction and valence bands move symmetrically with respect to the neutrality point.
The extended orbitals are only formed when the energies of the states localized in the n- and p-type dots are close to each other.  Figure \ref{np}(c) shows the charge density near the anticrossing of the localized energy levels
from the n-type and p-type dots.  The anticrossing indicates a presence of a tunnel coupling between the two quantum dots and a lack of any hidden symmetry difference
between the states of conduction and valence bands.

The orbitals in the n-p system change their character from ionic to extended as functions of the potential depth / height
with a 50\%/50\% distribution at the center of the avoided crossing. Note, that the avoided crossings for
each of the Kramers doublets is shifted one with respect to the other along the $V$ scale.

\subsection{Single-electron wave functions}

For the discussion of the two-electron interaction matrix elements, it is useful to look at the form of the single-electron wave functions.
For illustration  [Figure \ref{ff}] we consider a single n-type quantum dot and the $K'\uparrow$ state, i.e. the lowest-energy quantum-dot-confined state for $B>0$ (the center of the n-type quantum dot is set at $z=0$).
Figure \ref{ff}(a) shows the real part of the spin-up component with a rapid variation of the wave
function from ion to ion (blue and red colors correspond to opposite signs). The spatial variation of the wave functions in the nanotube can be put in an approximate form
\begin{equation}
 \Psi_{K^{(\prime)}\uparrow} =\exp(i{\bf K}^{(\prime)}\cdot {\bf r}+i\kappa^{(\prime)} R\theta) u({\bf r})\, ,\label{env}
\end{equation}
where $u$ is an envelope function, \cite{Bulaev} and for the zigzgag nanotube with 20 atoms along the circumference we have $\kappa'=(m-1/3)/R$ for $K$' valley
($\kappa=(m+1/3)/R$ for $K$ valley), where ${\bf K'}=(2\pi/a)(-1/3,1/\sqrt{3})$, ${\bf K}=(2\pi/a)(1/3,1/\sqrt{3})$,
$m$ is an integer, and $a=0.246$ nm. The nonzero value of $\kappa^{(\prime)}$ accounts for the amount that the wave vector satisfying the periodic boundary conditions misses the exact valley position \cite{minot}.

The lowest-energy confined states correspond to $m=0$ and Figs. \ref{ff}(b-c) show the real part of the envelope function $u({\bf r})$, i.e. the wave function $\Psi_{K'\uparrow}({\bf r})$ upon extraction of the rapidly varying valley factor $\exp(i{\bf K}'\cdot {\bf r}+i\kappa'R\theta)$.
The envelope $u({\bf r})$ is a smooth function separately on each of the nanotube sublattices A [Fig. \ref{ff}(b)] and B [Fig. \ref{ff}(c)].
We find that in the weak magnetic field and in the absence of the spin-orbit coupling, the  envelope $u$ is valley-independent. In presence of the spin-orbit coupling
the envelope function for the majority spin component is nearly the same for all
the four lowest-energy states independent of the spin-valley quantum numbers.
Some subtle differences can only be resolved for the avoided crossings
of the valence and the conduction band states [see Fig. \ref{np}(b) and the discussion below in Section \ref{fine}].
 Generally, for the majority spin components of the 4 low-energy states,
we have an approximate relation  $\Psi_{K}=\Psi_{K'} f_{KK'}$, with the $f_{KK'}$ factor rapidly varying in space that transforms the wave functions of $K'$ into $K$ valley,
$f_{KK'}=\exp(i({\bf K-K}')\cdot {\bf r}+i( \kappa -\kappa') R\theta)$.


\section{Two-electron states: The Method}

For the two-electron system we work with the energy operator including the electron-electron interaction,
\begin{equation}
 H_{2e}=\sum_a \epsilon_a g^\dagger_a g_a + \frac{1}{2} \sum_{abcd} V_{ab;cd}g^\dagger_a g^\dagger_b g_c g_d,\label{h2e}
\end{equation}
where $g^\dagger_a$ is the electron creation operator  in the eigenstate $a$ of the single-electron Hamiltonian, $\epsilon_a$ is the single-electron energy level,
and $V_{ab;cd}$ are the Coulomb matrix elements.
The Coulomb matrix elements are integrated in the real and spin space, as
\begin{equation} V_{ab;cd}=\langle \psi_{a}({\bf r_1},\boldsymbol{ \sigma}_1)\psi_{b}({\bf r_2},\boldsymbol{ \sigma}_2)|H_c |\psi_{c}({\bf r_1},\boldsymbol{ \sigma}_1)\psi_{d}({\bf r_2},\boldsymbol{ \sigma}_2)\rangle,\end{equation}
according to formula
\begin{widetext}
\begin{equation}
V_{ab;cd}=\sum_{i\sigma_i;j,\sigma_j;k,\sigma_k;l,\sigma_l}\alpha_{i,\sigma_i}^{a*}\alpha_{j,\sigma_j}^{b*}\alpha_{k,\sigma_k}^c\alpha_{l,\sigma_l}^d \delta_{\sigma_i;\sigma_k}\delta_{\sigma_j;\sigma_l}\langle p_z^i ({\bf r}_1) p_z^j ({\bf r_2}) |H_C| p_z^k ({\bf r_1}) p_z^l({\bf r_2}) \rangle,\end{equation} \end{widetext}
where $\alpha_{i,\sigma_i}^a$ is the contribution of $p_z^i$ orbital of spin $\sigma_i$ to the single-electron eigenstate $a$,
and $H_c$ is the Coulomb electron-electron interaction potential
\begin{equation}
H_C=\frac{e^2}{4\pi\epsilon\epsilon_0 r_{12}}
\end{equation}
with $r_{12} = |\boldsymbol{r_1}-\boldsymbol{r_2}|$.
We adopt the silicon dioxide dielectric constant $\epsilon=4$ as for the gated CNT coated in glass \cite{ambip}.
For calculation of the interaction matrix elements over the atomic orbitals we use the two-center approximation \cite{tc}: $\langle p_z^i p_z^j |\frac{1}{r_{ij}}| p_z^k p_z^l \rangle=\frac{1}{r_{ij}}\delta_{ik}\delta_{jl}$ for $i\neq j$. For the on-site integral ($i=j$) we take $\langle p_z^i p_z^j |H_C| p_z^i p_z^j \rangle=16.522$ eV  (after  Ref. \onlinecite{potasz}).

In the following for the n-n (p-p) system we set $V_l=V_r=-0.55$ eV ($+0.55$ eV) and for the n-p system $V_r=-V_l=0.42$ eV, unless stated otherwise.
We consider charging the energy levels which are the closest to the neutrality point.
For the n-n double dot in the summation over $a$ in Hamiltonian (\ref{h2e}) we include 8 energy levels of the bonding-antibonding pair, which correspond
to the energy of $\simeq -100$ meV at the vertical green line in Fig. \ref{wgd}(a) and additionally a number of higher-energy levels
(the number necessary for convergence depends on the size of the dot).
For the n-p double dot we consider the pair of energy levels of the avoided crossing marked by the green rectangle of Fig. \ref{np}(a)
at the avoided crossing of the conduction and valence bands and a number of higher-energy levels.
We assume that all the energy levels below are filled by electrons.
The higher-energy single-electron states introduce additional Slater determinants to the configuration-interaction basis.
Their contribution for the short quantum dots ($2d=4.4$ nm) is small,
and reliable results are obtained already for bases including 8 single-electron lowest-energy levels only.
However, a significant -- also qualitatively -- contribution of higher multiplets is present for larger quantum dots ($2d=30$ nm) that are considered in Section \ref{sectionlarge}.
Section \ref{sectionlarge} includes also the discussion of the convergence of the results.

\begin{figure}[htbp]
\includegraphics[width=8cm]{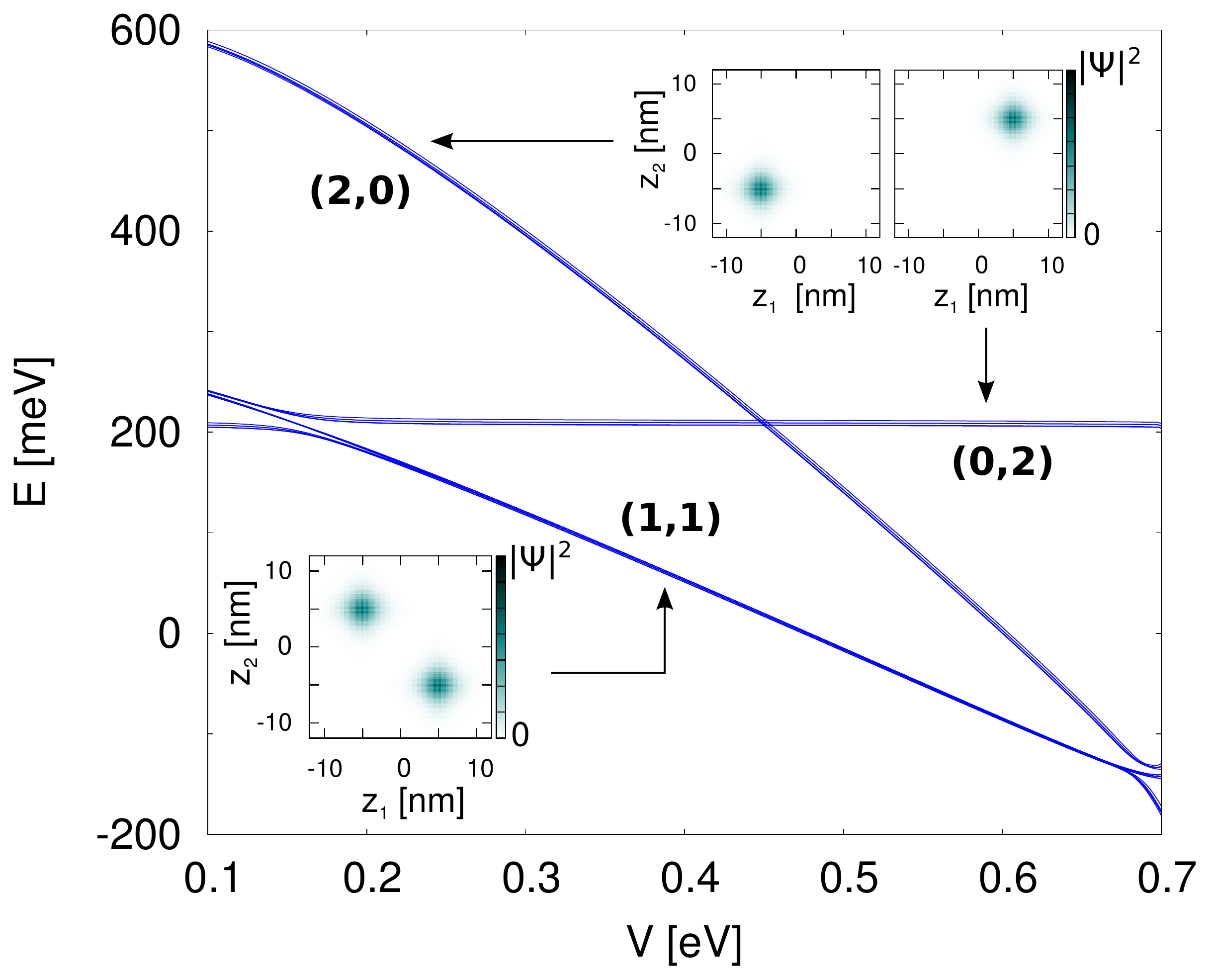}\\
\caption {Energy spectrum for the electron pair in n-p system as a function of the depth $V=-V_l$ of the n-dot. The p-dot potential is constant and set to $V_r=0.42$ eV. In the insets:
probability densities as functions of coordinates $z_1$ and $z_2$ (integrated over the CNT circumference) for both electrons.
There are 16 states for the electron distribution (1,1) and 6 states for configurations (2,0) and (0,2).}\label{asym2e}
\end{figure}

\begin{figure*}[htbp]

 \includegraphics[width=17.5cm]{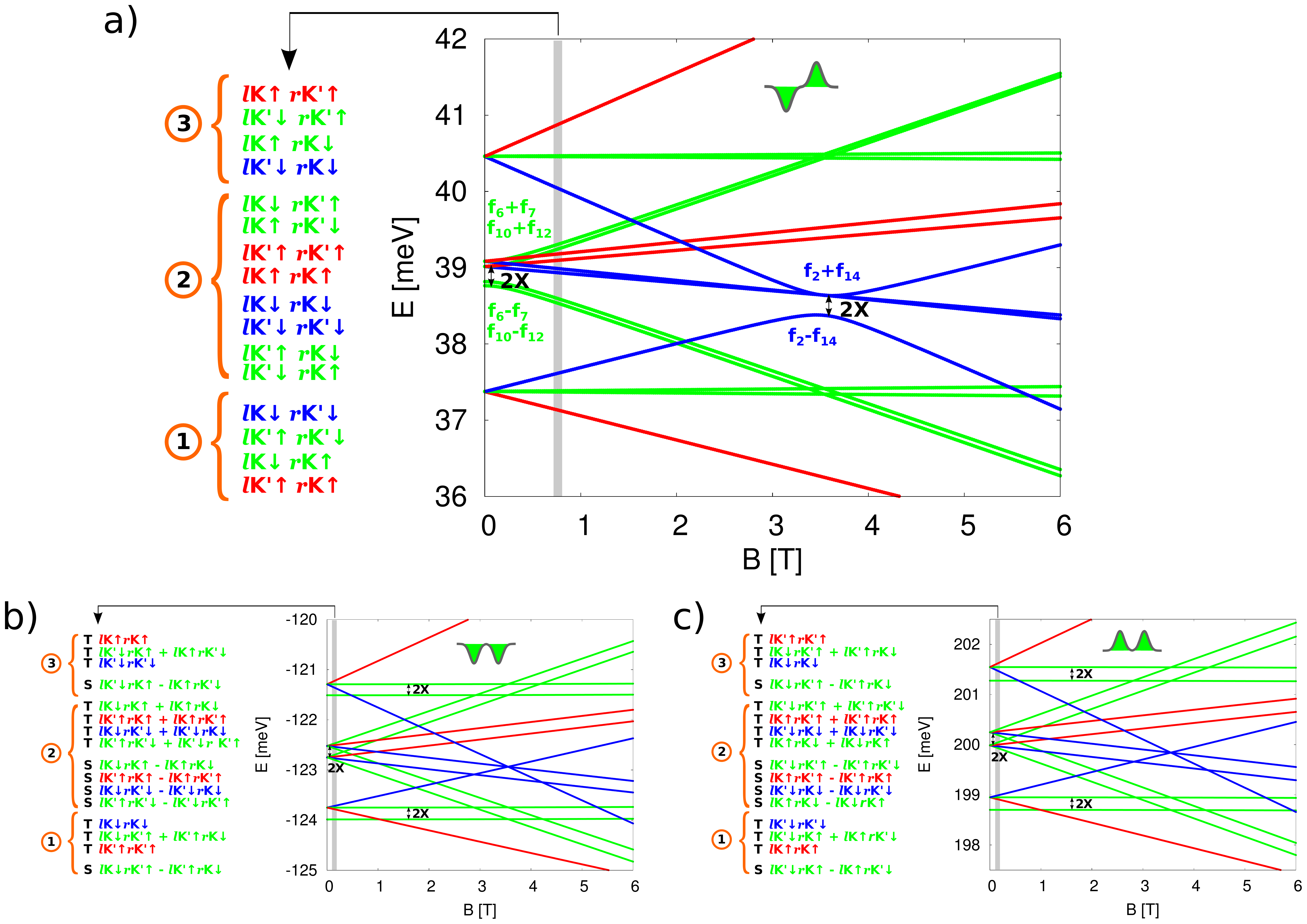}
\caption {Energy spectrum for the electron pair in the n-p system (a), a double n-dot (b), and double p-dot (c) as a function of the magnetic field $B$.
With the red (blue) color we plotted the energy levels of spin polarized up (down).
The green levels correspond to states of zero spin the $z$ component.  The integers 1, 2, and 3 number the group of energy levels. The single-electron energy levels
which contribute to these groups are explained in Fig. \ref{schemat2}. At the left of the plot we list the dominant configurations
that are found for a non-zero magnetic field (see the gray vertical belts).
In (b) and (c) we added labels S and T for singlet-like and triplet-like states
of spatial wave functions: symmetric and antisymmetric with respect to the electron interchange, respectively (see text).
In the avoided crossings opened by the exchange interaction in (a)
we denote the approximate form of the wave function as expressed with the Slater determinants $f_j$
that are listed in Table I.
Parameters of the system: distance $2z_s=10$ nm,  (a) $V_l=-V_r=-0.42$ eV, (b) $V_l=V_r=-0.55$  eV (c) $V_l=V_r=0.55$ eV.
}\label{2e}
\end{figure*}

\begin{figure}[htbp]
\includegraphics[width=8cm]{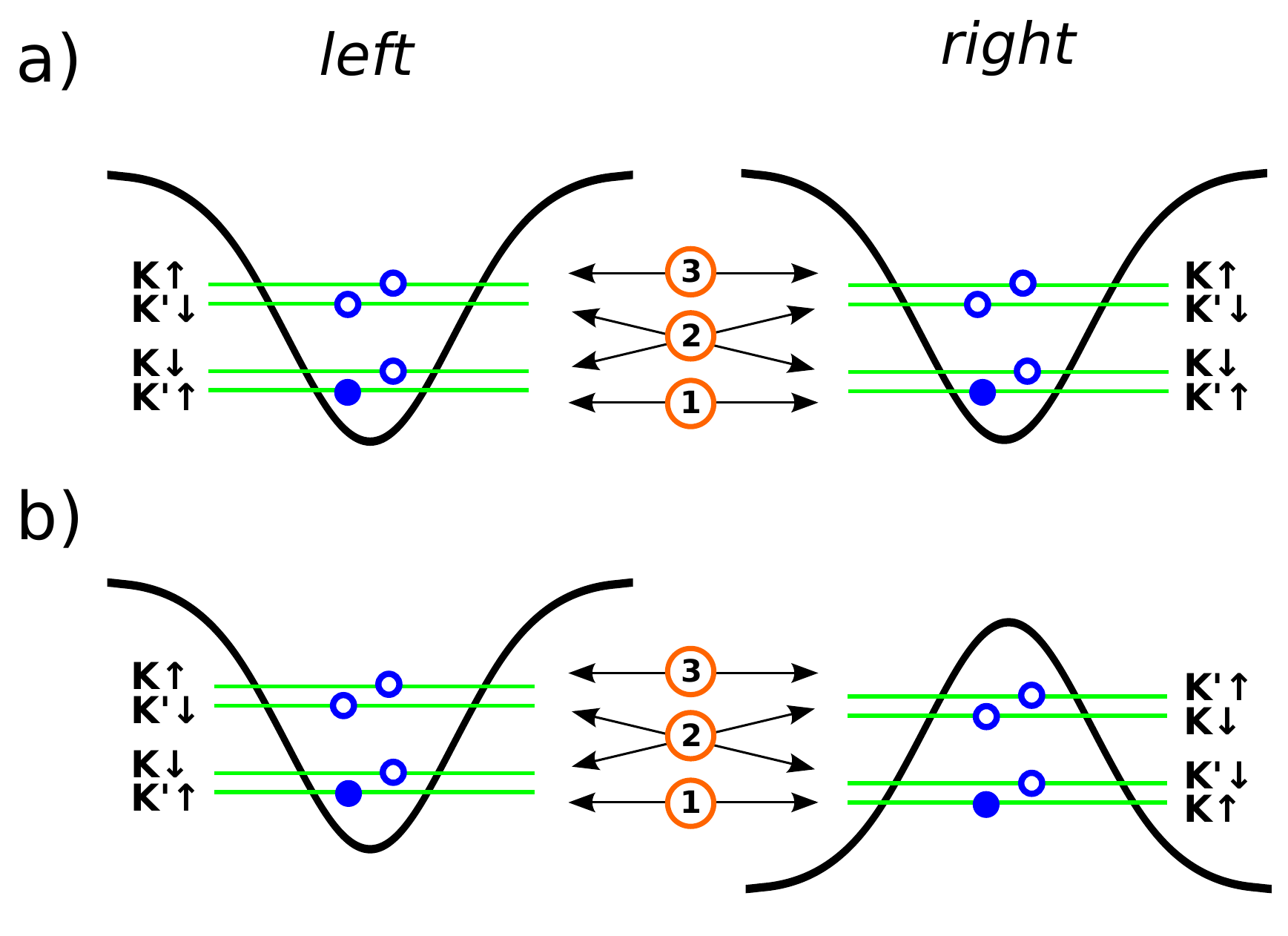}
\caption{Schematics of the two-electron systems considered in this paper in a double dot (a) and
in the n-p double dot (b). The filled (empty) circles correspond to occupied (unoccupied) single-electron orbitals. Valleys and spins of the single-electron energy levels split by the spin-orbit interaction are displayed. The arrows with labels 1, 2, 3 correspond to the dominant contributions to the two-electron energy levels that are discussed below. In (b) the localized states in the p-dot originate from the valence band, and only one of four accessible energy levels is occupied -- the configuration  corresponds to (1e,3h) charge state of the n-p double quantum dot.  }\label{schemat2}
\end{figure}

\section{Two electron states: Results}

In Fig. \ref{asym2e} we plotted the energy levels for the n-p double dot for $V_r=0.42$ eV as a function of $V_l=-V$.
The ground-state of the system in a wide range of $V_l$ corresponds to (1,1) electron distribution over the dots [or (1e,3h) according to notation of Ref. {\cite{pei,pei2}].
The system goes to the (0,2) ([0e,2h]) charge configuration at $V_l=-0.15$ eV and to (2,0) ([2e,0h]) at $V_l=-0.7$ eV.
The (1,1) energy level is nearly 16-fold degenerate, while (2,0) and (0,2) levels are 6-fold degenerate.

The 16 lowest-energy  two-electron states in the n-p, n-n, and p-p double dots
are displayed in  Fig. \ref{2e}(a,b) and (c), respectively.
The electrons in the 16 lowest-energy states occupy different dots [see the inset for (1,1) state in Fig. \ref{asym2e}], for the clarity of the discussion it is useful to consider the basis of single-electron states confined mostly in the left or right quantum dot.
The n-p double quantum dot is essentially asymmetric and the single-electron wave functions
exhibit a dominant localization in one of the dots [see Fig. \ref{np}(c)].
We  denote the states localized in the left and right dots as $l$ and $r$, respectively.
The adopted external potential of the n-n double quantum dot is symmetric and the electron
occupation of both the dots is 50\%-50\% in both the bonding and antibonding states.
In this case the $l$ and $r$ wave functions can be  constructed by a sum and a difference
of the bonding and antibonding wave functions.

The 16 lowest-energy two-electron levels at $B=0$ can be divided into three groups (see Figs. \ref{2e} and \ref{schemat2}).
The contributing basis elements for each of the groups are listed in Table I.
The four lowest-energy configurations that form the lowest energy levels at $B=0$ of Fig. \ref{2e}(a-c), is addressed as group '1' in Fig. \ref{2e}, Fig. \ref{schemat2}, and Table I. In this group the electron in each of the dots occupies one of the two-fold degenerate single-particle ground-states  (see Fig. \ref{schemat2}).

In Fig. \ref{2e} at the left-hand side of the plots we specify the dominant Slater determinant in the energy order that corresponds to the gray belt marked in the Figures \ref{2e}(a,b,c).
 We use the notation of Table I only with skipped antisymmetrization symbol.
The dominant Slater determinants for the two-electron states in the n-n and p-p systems differ by the inversion of valley indices ($K\leftrightarrow K'$).
All the systems -- including the n-p dot have an overall similar spin structure (see $S_z$ value as marked by colors in Fig. \ref{2e}).
The plots contain the lowest 16 energy levels for the (1,1) electron configuration.
In the n-n and p-p spectra there are 6 pairs of energy level  of the same component of the spin along the $z$ direction
which move parallel in $B$. The corresponding states differ in the symmetry of the two-electron spatial envelope which is either symmetric or antisymmetric  with respect to the electron interchange,
forming  the singlet-like and triplet-like states \cite{stecher,weiss,pei,pei2}.
The energy difference between energy levels of each couple is determined by the exchange energy, which remains essentially unchanged by $B$.
The corresponding pairs of energy levels for the n-p system are nearly degenerate [Fig. \ref{2e}(a)].
The n-n and p-p systems [Fig. \ref{2e}(b-c)] at $B=0$ have a non-degenerate ground state
and a three-fold degenerate excited state -- as in the single-triplet structure of  III-V double dots \cite{stecher,weiss,loss}.
On the other hand for the n-p double dot [Fig. \ref{2e}(a)] we find a four-fold degenerate ground-state which indicates a vanishing exchange energy.


\subsection{Exchange energy in the n-n system}

The  spin-orbit coupling in  CNTs  changes the energies of the states depending on the relative
orientation of the spin and angular momentum and introduces only a small contribution of the minority spin to the eigenstates.
Therefore, in the following analysis we refer to the majority spin component, only.
Let us consider $e_1$, $e_2$, $e_3$ and $e_4$ basis elements of Table I forming the lowest
energy group of energy levels denoted by (1) in Fig. \ref{2e}(b) and Fig. \ref{schemat2}.
For the spin polarized $e_1$ and $e_2$ basis elements   the spin-valley degree of freedom is separable from the spatial envelope, which is triplet-like, i.e. antisymmetric with respect to the electron interchange
\begin{equation}
e_1=\frac{1}{\sqrt{2}}(l(1)r(2)-r(1)l(2))K'\uparrow(1)K'\uparrow(2)\, ,
\end{equation}
and
\begin{equation}
\langle e_1 | H_C |e_1\rangle=\langle e_2 | H_C |e_2\rangle=C+X\, ,
\end{equation}
where $C$ is the Coulomb integral
\begin{equation}
 C=\langle l(1)r(2)|H_C|l(1)r(2)\rangle \, ,
\end{equation}
and $X>0$ is the exchange integral,
\begin{equation}
 X=-\langle l(1)r(2)|H_C|r(1)l(2)\rangle \,. \label{xc}
\end{equation}
The singlet-like energy levels are shifted down on the energy scale with respect
to the triplet-like energy levels by the exchange energy ($2X$) which is nearly independent of the
magnetic field [see Fig. \ref{2e}(b)].
The interaction integrals for the parameters of Fig. \ref{2e}
are  $C=38.75$ meV for the Coulomb and $2X=0.22$ meV for the exchange energy.

\begin{table}[!t]
\vspace{0.7cm}
\begin{tabular}{cccc}
$i$ & $group$ &n-n dot $e_i$ & n-p dot $f_i$ \\
\hline
$1$ & $ 1$ & ${\cal{A}}(l_{K'}^{\uparrow}(1)r_{K'}^\uparrow(2))$ & ${\cal{A}}(l_{K'}^\uparrow(1)r_K^\uparrow (2))$ \\
$2$ & $ 1$ & ${\cal{A}}(l_{K}^{\downarrow}(1)r_{K}^\downarrow(2))$ & ${\cal{A}}(l_K^\downarrow(1)r_{K'}^\downarrow (2))$ \\
$3$ & $ 1$ & ${\cal{A}}(l_{K}^{\downarrow}(1)r_{K'}^\uparrow(2))$ & ${\cal{A}}(l_K^\downarrow(1)r_K^\uparrow (2))$ \\
$4$ & $ 1$ & ${\cal{A}}(l_{K'}^\uparrow(1)r_{K}^\downarrow(2))$ & ${\cal{A}}(l_{K'}^\uparrow(1)r_{K'}^\downarrow (2))$ \\
\hline
$5$ & $ 2$ & ${\cal{A}}(l_{K'}^\uparrow(1)r_{K}^\uparrow(2))$ & ${\cal{A}}(l_{K'}^\uparrow(1)r_{K'}^\uparrow (2))$ \\
$6$ & $ 2$ & ${\cal{A}}(l_{K'}^\uparrow(1)r_{K'}^\downarrow(2))$ & ${\cal{A}}(l_{K'}^\uparrow(1)r_{K}^\downarrow (2))$ \\
$7$ & $ 2$ & ${\cal{A}}(l_{K}^\downarrow(1)r_{K}^\uparrow(2))$ & ${\cal{A}}(l_{K}^\downarrow(1)r_{K'}^\uparrow (2))$ \\
$8$ & $ 2$ & ${\cal{A}}(l_{K}^\downarrow(1)r_{K'}^\downarrow(2))$ & ${\cal{A}}(l_{K}^\downarrow(1)r_{K}^\downarrow (2))$ \\
$9$ & $ 2$ & ${\cal{A}}(l_{K}^\uparrow(1)r_{K'}^\uparrow(2))$ & ${\cal{A}}(l_{K}^\uparrow(1)r_{K}^\uparrow (2))$ \\
$10$ & $ 2$ & ${\cal{A}}(l_{K}^\uparrow(1)r_{K}^\downarrow(2))$ & ${\cal{A}}(l_{K}^\uparrow(1)r_{K'}^\downarrow (2))$ \\
$11$ & $ 2$ & ${\cal{A}}(l_{K'}^\downarrow(1)r_{K}^\downarrow(2))$ & ${\cal{A}}(l_{K'}^\downarrow(1)r_{K'}^\downarrow (2))$ \\
$12$ & $ 2$ & ${\cal{A}}(l_{K'}^\downarrow(1)r_{K'}^\uparrow(2))$ & ${\cal{A}}(l_{K'}^\downarrow(1)r_{K}^\uparrow (2))$ \\
\hline
$13$ & $ 3$ & ${\cal{A}}(l_{K}^{\uparrow}(1)r_{K}^\uparrow(2))$ & ${\cal{A}}(l_{K}^\uparrow(1)r_{K'}^\uparrow (2))$ \\
$14$ & $ 3$ & ${\cal{A}}(l_{K'}^{\downarrow}(1)r_{K'}^\downarrow(2))$ & ${\cal{A}}(l_{K'}^\downarrow(1)r_{K}^\downarrow (2))$ \\
$15$ & $ 3$ & ${\cal{A}}(l_{K'}^{\downarrow}(1)r_{K}^\uparrow(2))$ & ${\cal{A}}(l_{K'}^\downarrow(1)r_{K'}^\uparrow (2))$ \\
$16$ & $ 3$ & ${\cal{A}}(l_{K}^\uparrow(1)r_{K'}^\downarrow(2))$ & ${\cal{A}}(l_{K}^\uparrow(1)r_{K}^\downarrow (2))$ \\
\end{tabular}
\caption{16 lowest-energy Slater determinants basis elements for the n-n double dot  ($e_i$) and n-p double dot ($f_i$) with electrons occupying
separate quantum dots. ${\cal{A}}$ is the antisymmetrization operator with normalization factor $1/\sqrt{2}$,  $l$ / $r$
stand for the state localized in the left/right quantum dot, and  $(1)$, $(2)$ stand for the coordinates of the first and second electron respectively.
The numbers in the second column indicate the group of energy levels the determinant contribute to -- see Fig. \ref{2e} and \ref{schemat2}.
}
\end{table}

In the two-electron basis $e_3$ and $e_4$
with zero spin component in the $z$ direction ($S_z=0$)
 one cannot separate the spin-valley from the spatial coordinates in a similar manner.
The Coulomb interaction mixes the $e_3$ and $e_4$ configurations.
The diagonal interaction element for the third and fourth basis elements are
\begin{multline}
\langle e_3|H_C| e_3\rangle= \langle l_K^\downarrow(1)r_{K'}^\uparrow(2)| H_C| l_K^\downarrow(1)r_{K'}^\uparrow(2) \rangle \\
=C=\langle e_4|H_C| e_4\rangle\, ,
\end{multline}
and the non-diagonal

\begin{multline}
\langle e_3|H_C| e_4\rangle =-\langle l_K^\downarrow(1)r_{K'}^\uparrow(2)| H_C |r_{K}^\downarrow(1)l_{K'}^\uparrow(2)\rangle \\
=-\langle l(1)r(2)|H_C |r(1)l (2)\rangle=X. \,
\end{multline}
As a result we have a 2 by 2 Hamiltonian matrix
\begin{equation}
H_{XC}=\left(\begin{array}{cc}C&X \\ X& C\end{array}\right)
\end{equation}
with the energy eigenvalue $C-X$ for the singlet-like ground-state $s_{34}=e_3-e_4$
and $C+X$ for the excited triplet-like eigenstate $t_{34}=e_3+e_4$. The latter is degenerate with $e_1$ and $e_2$.
The singlet-like ground-state wave function is of the form
\begin{widetext}
\begin{equation}s_{34}=\frac{1}{2} (lK\downarrow(1) rK'\uparrow(2)-rK'\uparrow(1)lK\downarrow(2)-lK'\uparrow(1) rK\downarrow(2)+rK\downarrow(1)lK'\uparrow).\end{equation}
Upon replacement  $K=K'f_{KK'}$, one obtains \begin{equation} s_{34}={K'(1)K'(2)}\left[l(1)r(2)+r(1)l(2)\right]\left[f_{KK'}(1)\downarrow(1)\uparrow(2)-\uparrow(1)\downarrow(2)f_{KK'}(2)\right],\label{s34}\end{equation}
and similarly \begin{equation}t_{34}={K'(1)K'(2)}\left[l(1)r(2)-r(1)l(2)\right]\left[f_{KK'}(1)\downarrow(1)\uparrow(2)+\uparrow(1)\downarrow(2)f_{KK'}(2)\right].\label{t34}\end{equation}\end{widetext}
In $s_{34}$ and $t_{34}$ states the spin and valley are non-separable -- due to the presence of the intervalley scattering term $f_{KK'}$ in the spin-part of the formulae.
Nevertheless, the spatial wave function separates from the spin-valley and  has a  definite symmetry with respect to the electron interchange:  symmetric for $s_{34}$ (singlet-like state) and antisymmetric for $t_{34}$ (triplet-like state) -- see the first bracket in Eqs. \eqref{s34} and \eqref{t34}.

For the two-electron states of the other two groups of energy levels ("2" and "3" in Table I) the mixing of basis elements by the electron-electron interaction
occurs in a similar manner. In the spectrum one finds 6 pairs of two-electron energy levels that preserve their energy spacing
by $2X$ when $B$ is varied.

\subsection{The n-p system}

The lowest-energy group of the two-electron energy levels "1" ($f_1$, $f_2$, $f_3$ and $f_4$ in Table I) corresponds to each of electrons occupying the single-electron ground state
in one of the dots (cf. Fig. \ref{schemat2}).
The spin-polarized elements $f_1$ and $f_2$ separate from the rest of the group as in the n-n double dot.
For the n-p double dot  the lowest-energy states of the left and right dot of the same spin correspond to opposite valleys [see Fig. \ref{schemat2}(b)].
 Using the $f_{KK'}$ intervalley scattering function,  $f_1$ can be written as
 \begin{equation}
f_1={K'\uparrow(1)K'\uparrow(2)}(l(1)r(2)f_{KK'}(2)-r(1)f_{KK'}(1)l(2))\, .
\end{equation}
The interaction energy for this state is approximately equal to the Coulomb integral $\langle f_1|H_C | f_2\rangle=C$, since the exchange integral
$\langle(l(1)r(2)f_{KK'}(2)|H_C|r(1)f_{KK'}(1)l(2)\rangle$ involves
valley scattering for each of the electrons and thus it is negligibly small \cite{rontani2}.
For the same reason the off-diagonal matrix element $\langle f_3|H_C|f_4\rangle$ vanishes, with the diagonal
matrix elements equal to $C$. We are thus left with the fourfold degeneracy of the ground state as in Fig. \ref{2e}(a).
In none of the four lowest-energy eigenstates one can separate the spatial part of the spin-valley part
and in consequence, no singlet-like or triplet-like states in terms of the spatial envelope are formed.

In Fig. \ref{2e}(a) one finds two avoided crossings  -- one at $B=0$ for the $S_z=0$ states   (green curves) and another below $4$T for the spin-up polarized
states (blue curves).
The avoided crossing near 3.5T involves the $f_2$ state (group "1") and $f_{14}$ (group "3" --- see Table I).
Both these basis elements have the same $(K\downarrow, K'\downarrow)$ spin-valley configuration.
The energy level corresponding to $f_{14}$ ($f_2$)
-- decreases (increases) with increasing $B$ -- in consistence with the behavior of the lowest single-electron energy levels of the n- and p-dots [Fig. \ref{wgd}].
The interaction matrix element is then $\langle f_2|H_C|f_{14}\rangle= -\langle l_K^\downarrow(1)r_{K'}^\downarrow(2)|H_c|r_K^\downarrow(1)l_{K'}^\downarrow(2)\rangle= X$.
Thus the avoided crossings between these energy levels appear as due to the exchange interaction -- which is for the n-p system activated only when the single-electron energies are set equal by the external magnetic field.
In this sense, the external magnetic field induces formation of singlet-like and triplet-like states within the avoided crossing of energy levels.

For the n-p system the two-electron energy levels of the central group (2)  [near 39 meV at $B=0$ -- see Fig. \ref{2e}(a)] move in pairs with $B$ as for the n-n system, but now the pairs are nearly degenerate
and not split by the exchange energy.
The  8 energy levels of group (2) correspond to an electron in the ground-state of one of the dots, and an electron
in the excited state of the other dot  [see Fig. \ref{schemat2}(b)].
The pair of spin-down basis elements $f_8$ and $f_{11}$ correspond to both electrons
in $K$ and $K'$ valleys, respectively. For this reason the interaction matrix
elements is negligibly small and no avoided crossing between the energy levels is observed near $B=0$.
In $f_8$ and $f_{11}$ the valley and the spin are the same for both electrons and the wave function has
a separable form
 \begin{equation}
  f_{11}=K'(1)K'(2) \downarrow(1)\downarrow(2) (l(1)r(2)-l(2)r(1))\, ,
 \end{equation}
and both the spin-down basis elements $f_8$, $f_{11}$ produce triplet-like states.
The diagonal interaction matrix element is $C+X$ for both these states.  Same applies for the spin-up polarized states $f_5$ and $f_9$.

The remaining four $S_z=0$ states of group (2) can be divided into pairs in which the electrons occupy the same combinations
of spin-valleys:
 $K'\uparrow, K\downarrow$
 for $(f_6,f_7)$
and
 $K'\downarrow, K\uparrow$
for
 $(f_{10},f_{12})$.
For each of the pairs the diagonal matrix elements is  $C$ and off-diagonal interaction matrix element is $X$.
We obtain two-singlet like states: $s_{67}=f_6-f_7$, $s_{10,12}=f_{10}-f_{12}$, of interaction energy $C-X$ with
\begin{multline}
s_{67}=K'(1)K'(2)[l(1)r(2)+r(2)l(1)]\\
\times[\uparrow(1)\downarrow(2)f_{KK'}(2)-\downarrow(1)f_{KK'}(1)\uparrow(2)]\,,
\end{multline}
and two triplet-like states $t_{67}=f_6+f_7$, $t_{10,12}=f_{10}+f_{12}$ with energy $C+X$.
For $f_6$ basis element - one electron occupies the conduction band $K'\uparrow$ energy level and the other electron the valence band $K\downarrow$ energy level
which both decrease in $B$ -- see Fig. \ref{wgd}. The energy for its partner $f_7$ -- with interchanged bands for a given spin-valley -- increases with $B$. For $B>0.5$ T
the difference of the single-electron energies lifts the effects of the exchange interaction
and the energy levels become linear functions of $B$.

For the n-p system the interaction energies are very similar to the n-n dots with $C=38.76$ meV  and $2X=0.25$ meV -- in spite of the difference in $|V_{l/r}|$ values. This similarity is characteristic
to coupling of small quantum dots only (see Section \ref{sectionlarge}).


\begin{figure}[htbp]
\includegraphics[width=8cm]{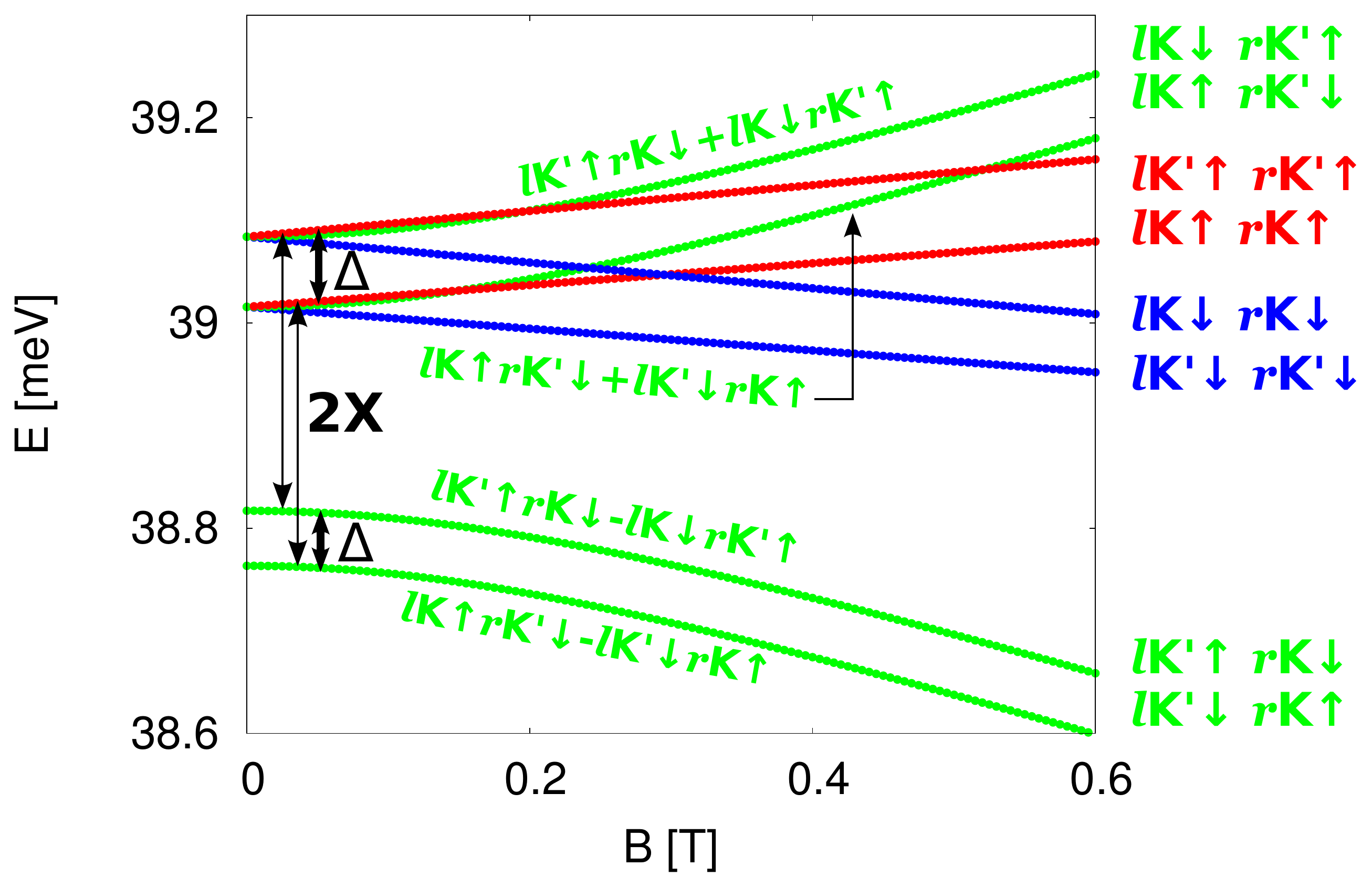}\\
\caption {A fragment of Fig. \ref{2e}(a) for the central group of energy levels (number 2). }\label{zoom}
\end{figure}

\subsection{Fine structure of the central level group at $B=0$ for the n-p system}
\label{fine}
According to the above discussion in the energy level group (2) at $B=0$ we should have a two-fold degenerate lower energy level of
singlet-like states and a fourfold degenerate triplet-like energy level.  In fact, we find [see a zoom in  Fig. \ref{zoom}] that
the energy levels are additionally split by an energy of $\Delta\simeq 0.06$  meV. 
This splitting  is {\it not} a result of the single-electron effects -- a difference in SO energy
splitting in the valence and conduction band for instance. In the present model the SO splitting energy is exactly the same in  both the dots.
 The fine structure is an interaction-mediated effect of the varied
 distribution of electrons within the n-p system.
 Let us look back at the avoided crossing of conduction and valence band energy levels of Fig. \ref{np}(c).
 The pair of nearly degenerate energy levels
 of the conduction and valence bands have inverted valley indices.
  The avoided crossing
 between the conduction- and valence-band states
for $K\downarrow, K'\uparrow$ spin-valley configuration appears for a lower value of $V$
 than for $K'\downarrow, K\uparrow$ states. Exactly at the center of each avoided crossings the electron distribution  within the n-p dot pair is 50\%/50\%.
 At $V=0.42$ eV for $K\downarrow$ and $K'\uparrow$  we are closer to the avoided crossing,
 and we find that each of the states of the n-p dot exhibits a slightly increased presence
of the probability density distribution in the other dot.
The difference is small, and so is the value of $\Delta$. The energy increase results from a larger electron-electron
interaction for $K\downarrow,K'\uparrow$ spin-valleys because of a less complete electron separation.
The avoided-crossing between the conduction and valence bands is the only case that we encountered when
the spatial localization depends on the spin-valley state.

The spin-polarized states in Fig. \ref{zoom} correspond to singlet-like and triplet-like spatial symmetry for any $B$.
On the other hand the $S_z=0$ states acquire a determined spatial symmetry with respect to the electrons interchange only
at the center of the avoided crossing ($B=0$) that is opened by the exchange interaction.

\begin{figure}[htbp]
\includegraphics[width=8cm]{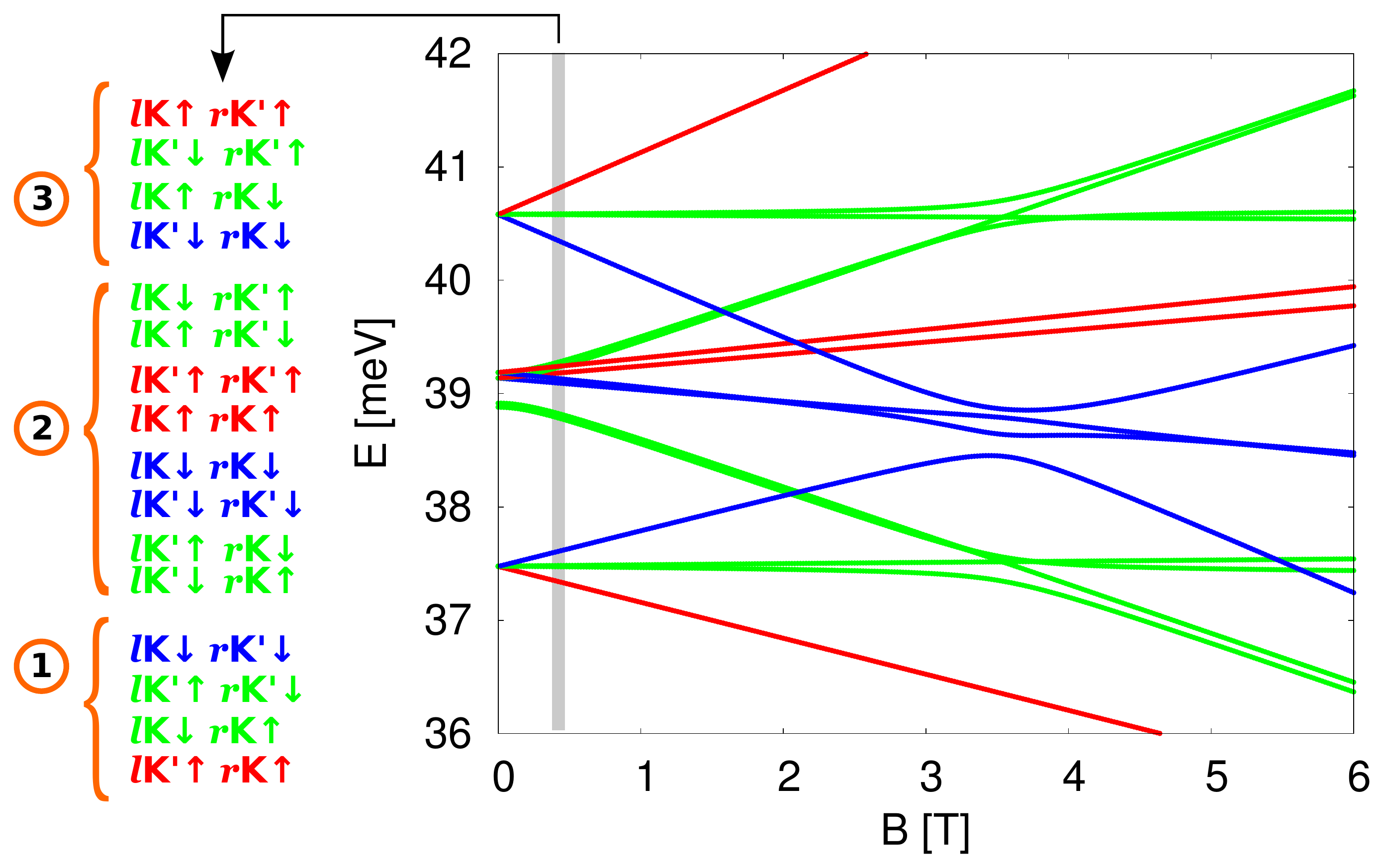}\\
\caption {Two-electron energy levels for the n-p dot in presence of the atomic disorder. An atom at  a distance of 8.5 nm to the left from the center of the system is removed. 
}\label{disorder}
\end{figure}

\subsection{Atomic disorder and valley mixing effects for the n-p spectrum}
The results presented so far were obtained for a clean CNT. In order to estimate the effect of the valley mixing induced by the lattice disorder
we removed one carbon atom at a distance of 8.5 nm to the left from the center of the system.
The results for the two-electron spectrum in the n-p dot are displayed in Fig. \ref{disorder}.
The valley mixing opens an avoided crossing near 3.5 T for the energy levels
that crossed near 3.2 T for a clean CNT [Fig. \ref{2e}(a)]. The crossing energy levels
corresponding to states ${\cal{A}} [lK'\downarrow(1)rK\uparrow(2)]$ and ${\cal{A}}[ lK\downarrow(1)rK\uparrow(2)]$ differ
by the valley index for one of the two electrons.
The lattice disorder induces  valley mixing and opens an avoided crossing between the corresponding energy levels of Fig. \ref{disorder}.
Outside these avoided crossings the spectrum resembles the one for a clean CNT [Fig. \ref{2e}(a)].
In particular, the near two-fold degeneracy of these energy levels
 -- in which {\it both} the electrons occupy different valleys [$(f_2,f_3), (f_5,f_6), (f_7,f_8), (f_9,f_{10}), (f_{14},f_{15})$ -- see Table I] is preserved also for $B\neq 0$.
The four-fold ground-state degeneracy at $B=0$ is not affected by the atomic disorder.

\subsection{Larger quantum dots}
\label{sectionlarge}

\begin{figure}[htbp]
\includegraphics[width=8.5cm]{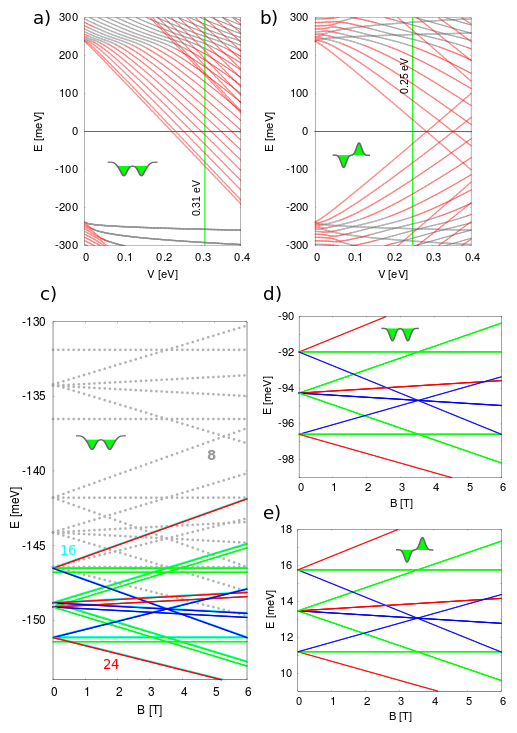}\\
\caption {Spectra for the system of the parameters $L=106.36$ nm, $2d=30$ nm, $z_s=12$ nm ($z_s=15$ nm for (d)).  (a)/(b) Energy levels for a system of n-n / n-p double dots as a function
of the depth $V=-V_l=-V_r$ / depth and height $-V_l=V_r=V$ of the Gaussian potential traps.  (c)/(e) Energy spectrum for the electron pair in the n-n / n-p system as a function of magnetic field $B$. In (c) the convergence of the results is shown with 8 (gray dots), 16 (light blue), and 24
single-electron states spanning the basis of the Slater determinants. The results for 24 basis
elements are given in the color palette for the spin-valleys as used in precedent figures.
(d) The same as (c) but for $z_s=15$ nm.
}\label{dots_12_15}
\end{figure}

In the experimental setups the quantum dots defined electrostatically in CNTs are longer,
and in consequence the single-electron energy level spacings are smaller than in the results presented above.
For longer quantum dots the contribution of higher single-electron spin-orbitals to the two-electron
states are more significant and the tunnel coupling for a fixed barrier width is reduced along with the
confinement energy.

In order to verify the conclusions reached for the model of small quantum dots we
performed calculations for the length of the dots increased from $2d=4.4$ nm to $2d=30$ nm,
which required dilatation of the nanotube from $L=53.1$ nm to $L=106.3$ nm.
The center of the dots were placed at a distance of $2z_s=24$ nm.
The results for the single-electron spectra are displayed in Fig. \ref{dots_12_15}(a,b),
with a pronounced reduction of the level spacing as compared to Fig. \ref{2kkd}(a) and Fig. \ref{np}(a).

The results for two-electrons in the n-n dot calculated for $V=0.31$ eV  are
displayed in Fig. \ref{dots_12_15}(c) for the basis of 8- (gray dotted lines),
16 (light blue curves) and 24 single-electron functions spanning the configuration-interaction basis of the Slater determinants. For each
choice of the basis we display 16 lowest-energy two-electron levels.
For 8 basis elements the 6 highest-energy levels (with energy above -137 meV) correspond to
the singlet-like states which climb up on the energy scale with respect to the triplet-like states.
 The variational overestimate for the singlet-like
states is much larger than for the 10 triplet-like states.
The slower convergence of the configuration-interaction method for  spin-singlets is
found also for III-V quantum dots \cite{szafran1}, and results from the fact
that for the spin triplets the antisymmetry of the spatial wave functions (Pauli exclusion)
keeps the electrons away, with the electron-electron correlation at least partly included
in the symmetry of the wave functions.  The results for 16 and 24 single-electron basis elements
are nearly identical, and the spectrum once the convergence is reached   is qualitatively the same as the one found for smaller quantum dots [cf. Fig. \ref{2e}(b), for 8 single-electron basis elements].
For the n-n dots with a larger interdot barrier [Fig. \ref{dots_12_15}(d) for $2z_s=30$ nm] the exchange energy becomes negligible. The spectrum for the n-p dot
displayed in Fig. \ref{dots_12_15}(d) exhibits no effects of the exchange interaction
already at $2z_s=24$ nm.

The exchange energy vanishes along with the overlap of the single-electron wave functions localized in both the dots (cf. Eq. (\ref{xc})).
As the size of the quantum dots increases, the tunnel coupling between the dots
disappears faster for the n-p system as compared to the unipolar n-n or p-p quantum dots.
For the n-p dot, the electron of the type-n dot needs to climb the potential hill defining the type-p dot
to form an extended state.
Note, that the experimental results of Fig. 1(c) of Ref. \onlinecite{pei} for the current
as a function of $V_l$ and $V_r$ voltages indeed demonstrate that lifting of the Coulomb blockade
for the unipolar dots appears for a wider range of  gate voltages than for the n-p dot, suggesting
a reduced tunnel coupling between the ambipolar  dots.

For a Gaussian profile of the confinement potential
the reduction of the  exchange energy for the n-p dots appears already for smaller dots
-- see Fig. \ref{posredni} for $2d=14$ nm and $2z_s=10$ nm, for which the exchange energy is $2X=0.1$ meV.
For larger dots the exchange energy in the n-p system appears when the the n-p junction is 
shorter. In Fig. \ref{sing} we present calculation for the confinement potential
of form
\begin{equation}
V=\left\{
\begin{array}{l l}
-V \exp(-(z+z_s)^2/d^2) & \quad \text{for $z<-z_s$}\\
V\sin{(\pi z/(2z_s))} & \quad \text{for $-z_s\leq z\leq z_s$}\\
V \exp(-(z-z_s)^2/d^2) & \quad \text{for $z>z_s$}
\end{array}
\right. \label{singe}
\end{equation}
for $V=0.23$ eV, $z_s=3$ nm, $d=20$ nm. This potential profile  is plotted in Fig. \ref{sing}(b) with the black line.
An overlap of the wave functions of both dots appear [see Fig. \ref{sing}(b)] near the center
of the system, and the exchange energy is again significant ($2X=0.22$ meV). Then, the two-electron energy spectrum
takes the form [Fig. \ref{sing}(c)] from the discussion of small quantum dots [Fig. \ref{2e}(a)].

\begin{figure}[htbp]
\includegraphics[width=6.5cm]{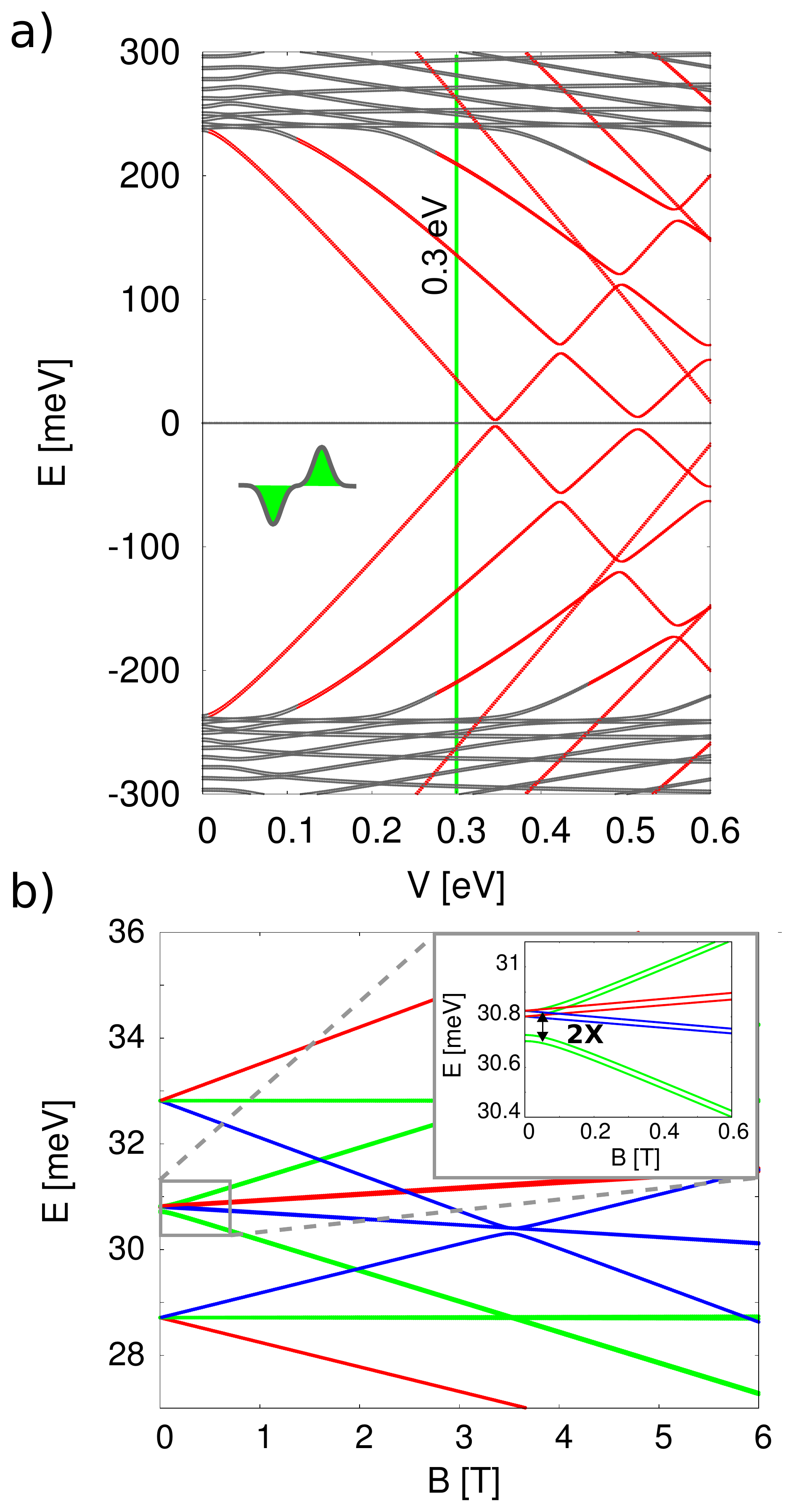}\\
\caption {(a) Single electron spectrum for n-p quantum dot with $2d=14$ nm, and $2z_s=10$ nm.
(b) The two-electron spectrum. The inset shows the zoom on the central part of the spectrum.
The colors stand for the spin-configuration with the palette of Fig. \ref{2e}.}\label{posredni}
\end{figure}

\begin{figure}[htbp]
\includegraphics[width=6.5cm]{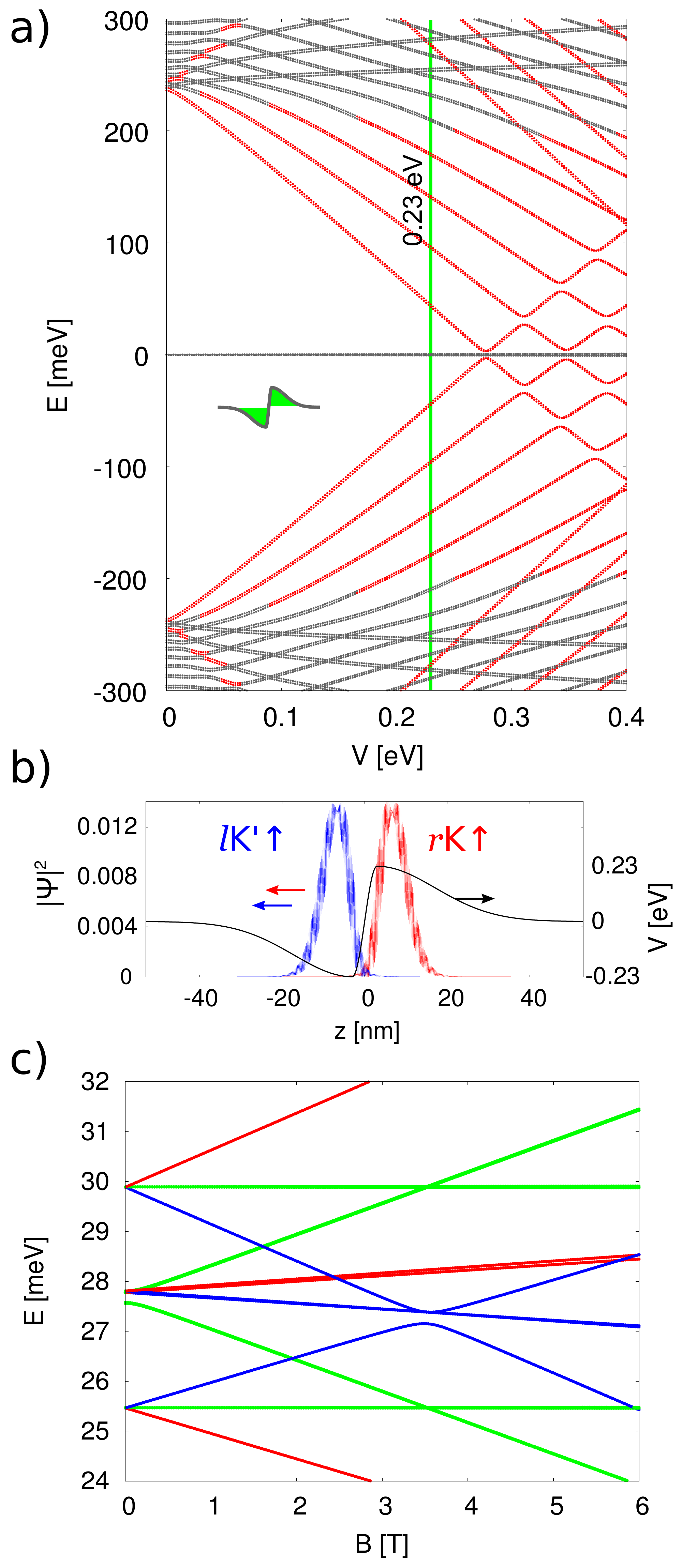}\\
\caption {Single- (a) and two- (c) electron energy spectra for the n-p quantum dot with
confinement potential given by Eq. \ref{singe} and plotted with the black line in (b).
In (b) the blue and red lines show the wave functions for the single-dot eigenstates.}\label{sing}
\end{figure}

\subsection{CNT chirality and the spin-orbit coupling parameter $\delta$}

The presented results are qualitatively independent of the chirality of the CNT,
as long as it is semiconducting. For presentation we  return to the parameters
of the small Gaussian quantum dots and consider a $C_h=(20,6)$ CNT [Fig. \ref{schematchiral}].
For $2z_s=10$ nm -- the distance between the centers of the dots for the zigzag CNT considered in Section \ref{smol} --
a wide avoided crossing is found in the single-electron states from the conduction and the valence bands
[Fig. \ref{chiral}(a)] and the exchange energy is as large as $2X=0.83$ meV (see Fig. \ref{chiral}(b)).
For $2z_s=11.26$ nm the width of the avoided crossing of the single-electron energy levels is reduced to 7.7 meV
[exactly as for the zigzag dot of Fig. \ref{np}(b)], and the exchange energy is $2X=0.25$ meV.
The qualitative character of the n-p spectrum, including the pattern of the avoided crossings
 is the same as for the zigzag CNT [cf. Fig. \ref{chiral}(b) and Fig. \ref{2e}(a)].

\label{chirals}
The spin-orbit interaction in the applied model is determined by the parameter $\delta$ [Eqs. (3,4)]. The sign of $\delta$ determines the
sign of the spin-orbit splitting $\Delta_{SO}$ between  ($K'\uparrow$, $K\downarrow$) and ($K'\downarrow$, $K\uparrow$) energy levels.
Depending on the sign of $\delta$ the energy levels of the multiplet cross as a function of the magnetic field
in the lower \cite{Bulaev} or higher \cite{soc1} pair of energy levels. Both types of crossings are observed in experiments for various samples \cite{str2}.
In order to demonstrate that the conclusions of the present study are independent of the sign of $\Delta_{SO}$ we performed
calculations for the small dots within the zigzag CNT adopting $\delta=-0.003$. The single-particle energy spectra change
are displayed in Fig. \ref{wgd_delta-} for a single carrier and in Fig. \ref{b2edg_delta-} for the n-p quantum dot.

\begin{figure}[htbp]
\includegraphics[width=6cm]{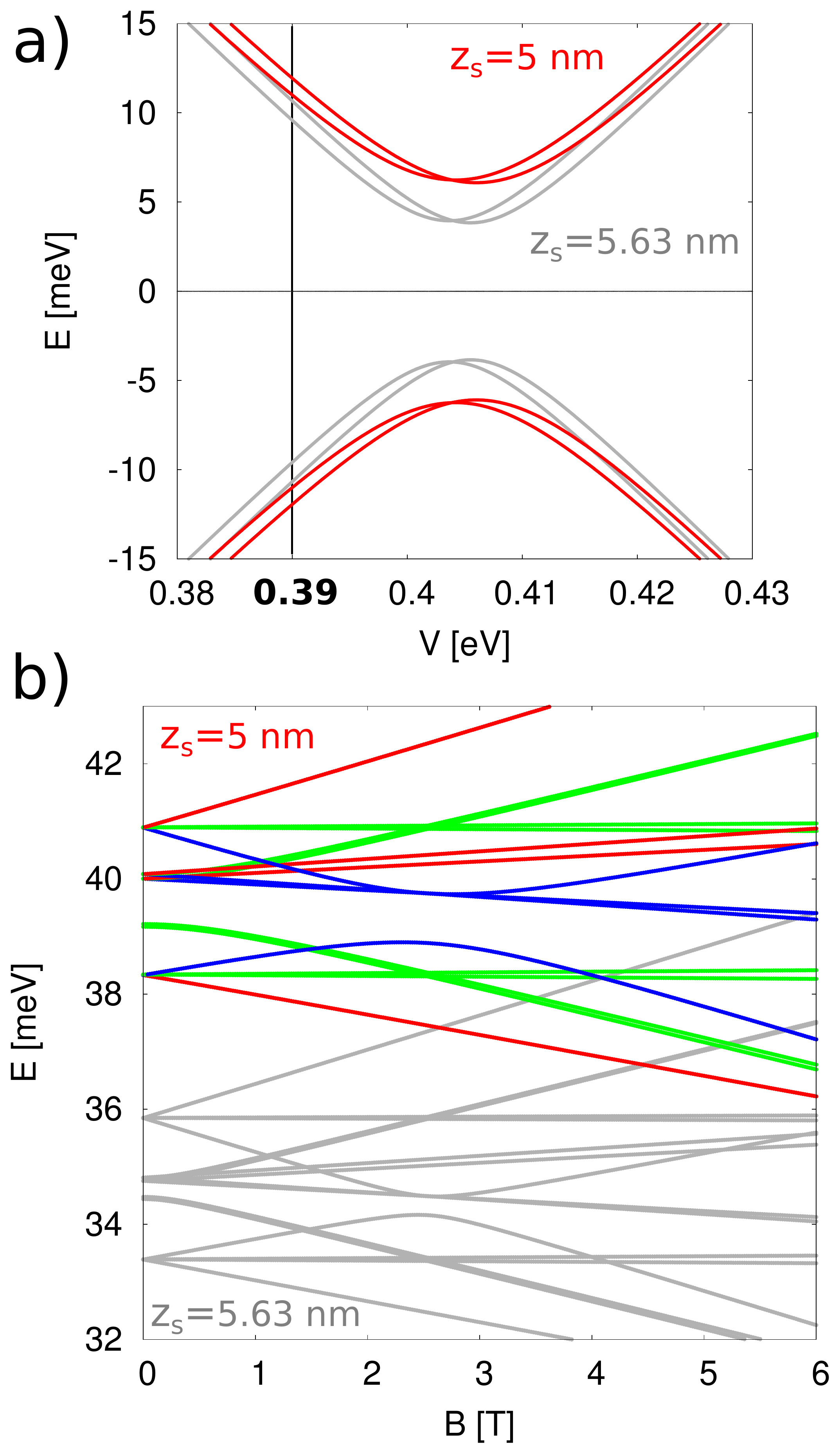}\\
\caption{ (a) Avoided crossing of valence and conduction band single-electron states for a n-p quantum dot within(20,6) CNT.
[The results for the zigzag CNT were displayed in Fig. \ref{np}(b)]. (b) Two-electron
energy spectrum. Two values of interdot separation are considered $2z_s=10$ nm and $2z_s=11.26$ nm.
}\label{chiral}
\end{figure}

\begin{figure}[htbp]
\includegraphics[width=6cm]{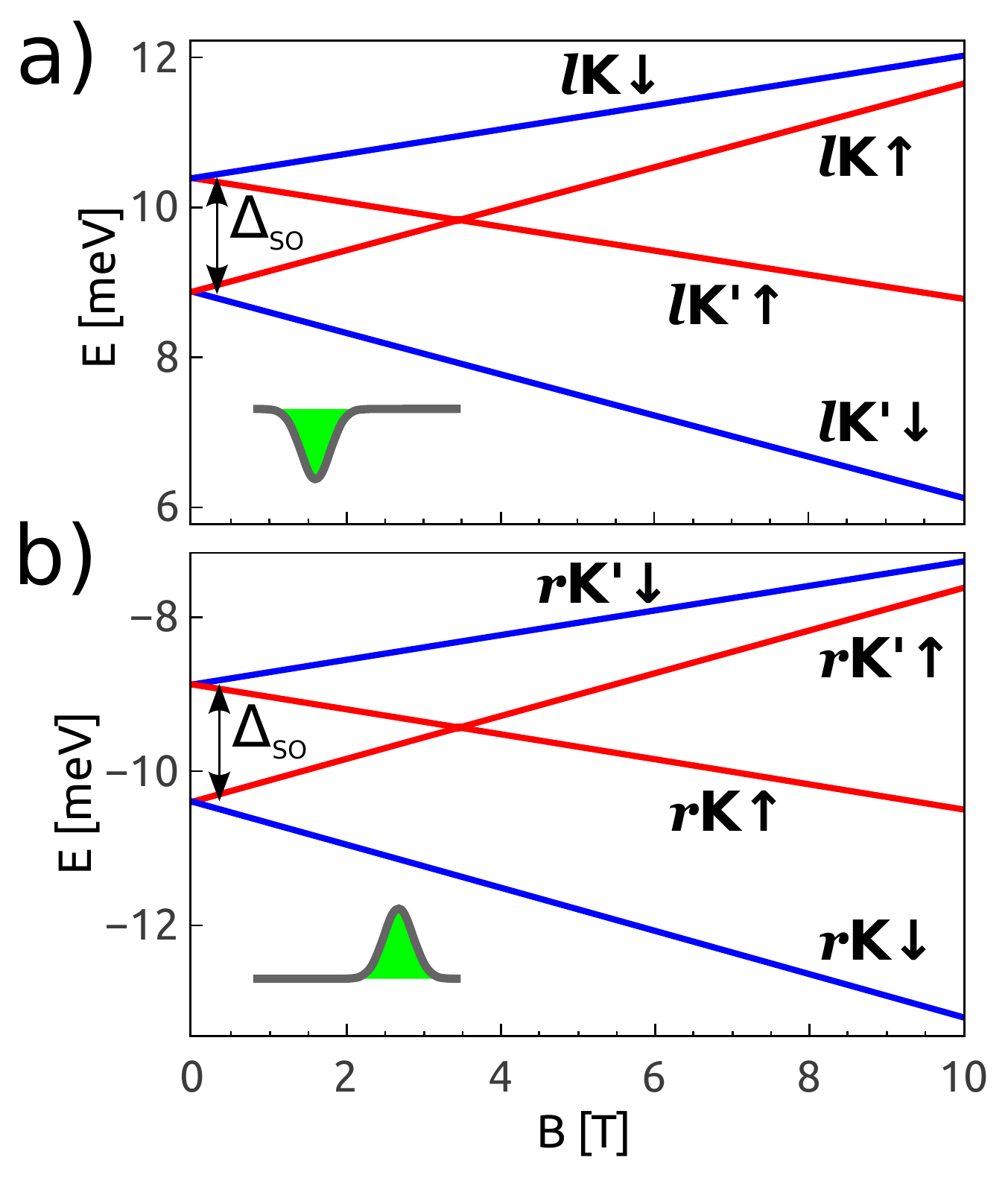}\\
\caption{The same as Fig. \ref{wgd}(a-b) but for $\delta=-0.003$.
}\label{wgd_delta-}
\end{figure}

\begin{figure}[htbp]
\includegraphics[width=8cm]{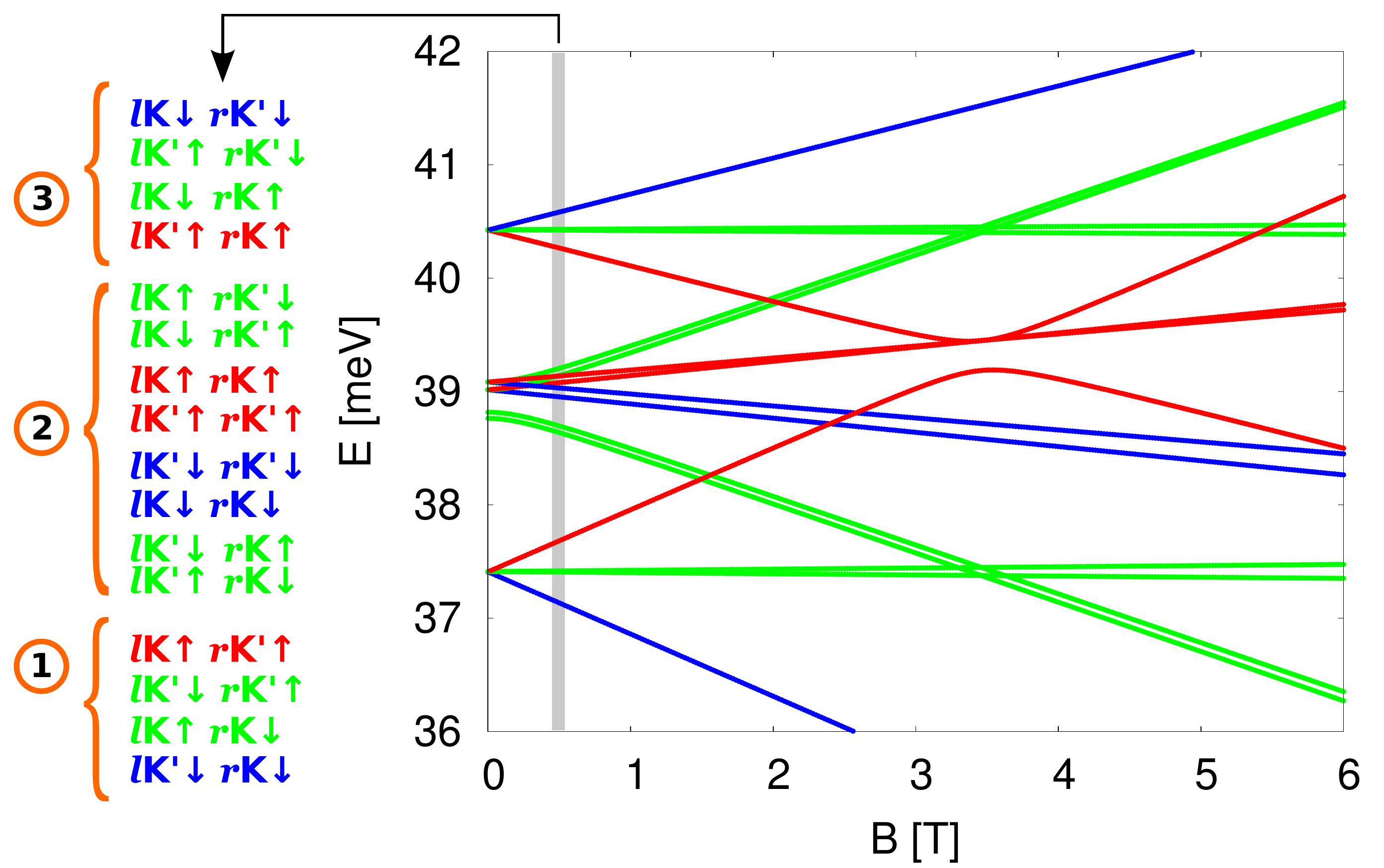}\\
\caption {The same as Fig. \ref{2e}(a) but for $\delta=-0.003$.
}\label{b2edg_delta-}
\end{figure}

For the negative value of $\delta$ the spin-orbit splitting favors parallel alignment of the orbital and the spin magnetic moments, in contrast to the results presented above for the positive $\delta$.
This leads to the switched order of the two Kramers doublets on the energy scale for $B=0$ (Fig.  \ref{wgd_delta-}). Furthermore, the crossing of the single electron states in positive magnetic field
appear now in higher pair of states -  $K\,(K')$ for n-type (p-type) dot - instead of lower [cf. Fig. \ref{wgd}(a-b)].
Changes in the single electron spectra are projected directly to the electron-pair spectrum - Fig. \ref{b2edg_delta-}.
Comparing Fig. \ref{b2edg_delta-} to Fig. \ref{2e}(a) we conclude that the avoided crossing in the central part of the spectrum for $B\simeq 3.5$T is
observed either for the spin-down states [Fig. \ref{2e}(a)] or spin-up states [Fig. \ref{b2edg_delta-}] depending on the sign of $\delta$.

\section{Summary and Conclusion}

We have described formation of extended single-electron orbitals in n-p quantum dots defined in a carbon nanotube
and the two electron states corresponding to (1e,3h) charge state of the double dot.
The electronic structure was determined by the configuration interaction approach within the tight-binding method
with a complete account for the intervalley scattering due to the atomic disorder and electron-electron interaction
without any additional parameters describing the coupling of the conduction and valence band states.

The present study indicates that the exchange energy for the n-p dots appears only for finite intervals of the magnetic field and only in some parts of the spectrum.
In particular, the spin exchange interaction is missing in the ground-state, which is fourfold degenerate at $B=0$.
The reason for this unusual behavior of the exchange interaction -- as compared to n-n quantum dots -- is the fact that for a given valley
the orbital momenta are opposite in the conduction and valence bands. Formation of singlet-like and
triplet-like orbitals appears only briefly on the $B$ scale and the ground-state is four-fold degenerate.
For a general value of $B$ the exchange integral
vanishes by the valley orthogonality.
The basic structure of the two-electron spectrum turns out to be robust against the atomic disorder, chirality, the sign of $\Delta_{SO}$
and the size of the dots -- provided that a tunnel coupling between the quantum dots is present.
The tunnel coupling for the n-p dots is generally more difficult to obtain than for the unipolar dots
and requires a short n-p junction to allow for the overlap of the single-dot wave functions.
 The present study indicates
that the ground state of the two-electron n-p dot is four-fold degenerate also when the n-p dots are strongly coupled.
\section*{Acknowledgements}
This work was supported by National Science Centre
according to decision DEC-2013/11/B/ST3/03837 and by
PL-GRID infrastructure.


\begin{thebibliography}{1}
\bibitem{uwaga} We identify the valleys by inspection of the angular wave function dependends after Ref. \cite{Bulaev}
as described in Ref. \cite{eno}.
\end{thebibliography}
\end{document}